%% file: main.tex
\pdfoutput=1
\documentclass[12pt,a4paper]{article}

\usepackage{ifthen} 
\newboolean{pdflatex}
\setboolean{pdflatex}{true} 

\newboolean{articletitles}
\setboolean{articletitles}{true} 

\newboolean{uprightparticles}
\setboolean{uprightparticles}{false} 

\def\paperauthors{LHCb collaboration} 
\def\paperasciititle{Template for writing LHCb papers} 
\def\papertitle{Measurement of branching fractions and $C\!P$ asymmetries in $\mathit{\Lambda}_b^0(\mathit{\Xi}_b^0)\!\to pK_{\mathrm S}^0h^-$ decays} 
\def\paperkeywords{{High Energy Physics}, {LHCb}} 
\def\papercopyright{\the\year\ CERN for the benefit of the LHCb collaboration} 
\def\paperlicence{CC BY 4.0 licence}
\def\paperlicenceurl{https://creativecommons.org/licenses/by/4.0/}

\newif\ifEnableSectionTOCLinks
\EnableSectionTOCLinksfalse 

\input{preamble}
\usepackage{longtable} 

\input{my-symbols-def}

\begin{document}

\renewcommand{\thefootnote}{\fnsymbol{footnote}}
\setcounter{footnote}{1}

\input{title-LHCb-PAPER}


\renewcommand{\thefootnote}{\arabic{footnote}}
\setcounter{footnote}{0}


\cleardoublepage


\pagestyle{plain} 
\setcounter{page}{1}
\pagenumbering{arabic}


\input{body}

\input{acknowledgements}




\addcontentsline{toc}{section}{References}
\bibliographystyle{LHCb}
\bibliography{main,standard,LHCb-PAPER,LHCb-CONF,LHCb-DP,LHCb-TDR}

\newpage
\input{Authorship_LHCb-PAPER-2025-016}

\end{document}

%% file: preamble.tex
\usepackage[top=1in, bottom=1.25in, left=1in, right=1in]{geometry}

\usepackage{microtype}
\usepackage{lineno}  
\usepackage{xspace} 
\usepackage{caption} 

\usepackage{graphicx}  
\usepackage{color}
\usepackage{colortbl}
\graphicspath{{./figs/}} 

\usepackage{amsmath} 
\usepackage{amssymb}
\usepackage{amsfonts}
\usepackage{upgreek} 

\newcommand*\patchAmsMathEnvironmentForLineno[1]{%
\expandafter\let\csname old#1\expandafter\endcsname\csname #1\endcsname
\expandafter\let\csname oldend#1\expandafter\endcsname\csname
end#1\endcsname
 \renewenvironment{#1}%
   {\linenomath\csname old#1\endcsname}%
   {\csname oldend#1\endcsname\endlinenomath}%
}
\newcommand*\patchBothAmsMathEnvironmentsForLineno[1]{%
  \patchAmsMathEnvironmentForLineno{#1}%
  \patchAmsMathEnvironmentForLineno{#1*}%
}
\AtBeginDocument{%
\patchBothAmsMathEnvironmentsForLineno{equation}%
\patchBothAmsMathEnvironmentsForLineno{align}%
\patchBothAmsMathEnvironmentsForLineno{flalign}%
\patchBothAmsMathEnvironmentsForLineno{alignat}%
\patchBothAmsMathEnvironmentsForLineno{gather}%
\patchBothAmsMathEnvironmentsForLineno{multline}%
\patchBothAmsMathEnvironmentsForLineno{eqnarray}%
}

\usepackage[pdftex,
            pdfauthor={\paperauthors},
            pdftitle={\paperasciititle},
            pdfkeywords={\paperkeywords}]{hyperref}
\usepackage{hyperxmp}
\hypersetup{
    pdfcopyright={Copyright (C) \papercopyright},
    pdflicenseurl={\paperlicenceurl}
}

\usepackage[colorinlistoftodos,textsize=scriptsize]{todonotes}

\usepackage[bottom,flushmargin,hang,multiple]{footmisc}

\usepackage[all]{hypcap} 

\input{lhcb-symbols-def} 

\hypersetup{
  colorlinks   = true, 
  urlcolor     = blue, 
  linkcolor    = blue, 
  citecolor    = red   
}

\ifEnableSectionTOCLinks
    \usepackage[explicit]{titlesec} 
    
    \let\oldcontentsline\contentsline
    \renewcommand

    \titleformat{\section}{\normalfont\Large\bf}{\hyperlink{tocsection.\thesection}{{\thesection} \parbox[t]{\dimexpr\textwidth-1pc}{#1}}}{1pc}{}

    \titleformat{\subsection}{\normalfont\bf}{\hyperlink{tocsubsection.\thesubsection}{{\thesubsection} \parbox[t]{\dimexpr\textwidth-1pc}{#1}}}{1pc}{}

    \titleformat{name=\section,numberless}[display]{}{}{0pt}{\normalfont\Huge\bfseries #1}
\fi

\usepackage{cite} 
\usepackage{mciteplus}

%% file: lhcb-symbols-def.tex
\usepackage{xspace} 
\usepackage{upgreek}

\def\lhcb   {\mbox{LHCb}\xspace}

\def\MagUp {\mbox{\em Mag\kern -0.05em Up}\xspace}

\ifthenelse{\boolean{uprightparticles}}%
{

 \def\Pmu         {\ensuremath{\upmu}\xspace}                 
 \def\Pnu         {\ensuremath{\upnu}\xspace}                 
                  
 \def\Ppi         {\ensuremath{\uppi}\xspace}

 \def\Ppsi        {\ensuremath{\uppsi}\xspace}

 \def\PDelta      {\ensuremath{\Delta}\xspace}                 
 \def\PXi         {\ensuremath{\Xi}\xspace}                 
 \def\PLambda     {\ensuremath{\Lambda}\xspace}                 
 \def\PSigma      {\ensuremath{\Sigma}\xspace}                 
 \def\POmega      {\ensuremath{\Omega}\xspace}                 
 \def\PUpsilon    {\ensuremath{\Upsilon}\xspace}
 \let\oldPi\Pi
 \def\PPi         {\ensuremath{\oldPi}\xspace}

 \def\PB      {\ensuremath{\mathrm{B}}\xspace}                 
 \def\PD      {\ensuremath{\mathrm{D}}\xspace}                 
 \def\PJ      {\ensuremath{\mathrm{J}}\xspace}                 
 \def\PK      {\ensuremath{\mathrm{K}}\xspace}                 
 \def\Pb      {\ensuremath{\mathrm{b}}\xspace}                 
 \def\Pc      {\ensuremath{\mathrm{c}}\xspace}                 
 \def\Pd      {\ensuremath{\mathrm{d}}\xspace}                 

 \def\Pp      {\ensuremath{\mathrm{p}}\xspace}                 
 \def\Pq      {\ensuremath{\mathrm{q}}\xspace}                 
 \def\Ps      {\ensuremath{\mathrm{s}}\xspace}                 
                  
 \def\Pu      {\ensuremath{\mathrm{u}}\xspace}                 
 \def\thebaroffset{0.0em}
}
{

 \def\Pmu         {\ensuremath{\mu}\xspace}                 
 \def\Pnu         {\ensuremath{\nu}\xspace}                 
                  
 \def\Ppi         {\ensuremath{\pi}\xspace}

 \def\Ppsi        {\ensuremath{\psi}\xspace}                 
                  
 \mathchardef\PDelta="7101
 \mathchardef\PXi="7104
 \mathchardef\PLambda="7103
 \mathchardef\PSigma="7106
 \mathchardef\POmega="710A
 \mathchardef\PUpsilon="7107
 \mathchardef\PPi="7105
 \def\PB      {\ensuremath{B}\xspace}                 
 \def\PD      {\ensuremath{D}\xspace}                 
 \def\PJ      {\ensuremath{J}\xspace}                 
 \def\PK      {\ensuremath{K}\xspace}                 
 \def\Pb      {\ensuremath{b}\xspace}                 
 \def\Pc      {\ensuremath{c}\xspace}                 
 \def\Pd      {\ensuremath{d}\xspace}                 

 \def\Pp      {\ensuremath{p}\xspace}                 
 \def\Pq      {\ensuremath{q}\xspace}                 
 \def\Ps      {\ensuremath{s}\xspace}                 
                  
 \def\Pu      {\ensuremath{u}\xspace}                 
 \def\thebaroffset{0.18em}
}
\newcommand{\offsetoverline}[2][\thebaroffset]{\kern #1\overline{\kern -#1 #2}}%

\makeatletter
\ifcase \@ptsize \relax
  \newcommand{\miniscule}{\@setfontsize\miniscule{4}{5}}
\or
  \newcommand{\miniscule}{\@setfontsize\miniscule{5}{6}}
\or
  \newcommand{\miniscule}{\@setfontsize\miniscule{5}{6}}
\fi
\makeatother

\DeclareRobustCommand{\optbar}[1]{\shortstack{{\miniscule (\rule[.5ex]{1.25em}{.18mm})}
  \\ [-.7ex] $#1$}}

\def\mup        {{\ensuremath{\Pmu^+}}\xspace}
\def\mun        {{\ensuremath{\Pmu^-}}\xspace} 

\def\neub       {{\ensuremath{\overline{\Pnu}}}\xspace}

\def\neumb      {{\ensuremath{\neub_\mu}}\xspace}

\def\quark     {{\ensuremath{\Pq}}\xspace}

\def\uquark    {{\ensuremath{\Pu}}\xspace}
\def\uquarkbar {{\ensuremath{\overline \uquark}}\xspace}

\def\dquark    {{\ensuremath{\Pd}}\xspace}

\def\squark    {{\ensuremath{\Ps}}\xspace}

\def\cquark    {{\ensuremath{\Pc}}\xspace}

\def\bquark    {{\ensuremath{\Pb}}\xspace}

\def\pion   {{\ensuremath{\Ppi}}\xspace}
\def\piz    {{\ensuremath{\pion^0}}\xspace}
\def\pip    {{\ensuremath{\pion^+}}\xspace}
\def\pim    {{\ensuremath{\pion^-}}\xspace}

\def\kaon    {{\ensuremath{\PK}}\xspace}

\def\KorKbar {\kern \thebaroffset\optbar{\kern -\thebaroffset \PK}{}\xspace}

\def\Kp      {{\ensuremath{\kaon^+}}\xspace}
\def\Km      {{\ensuremath{\kaon^-}}\xspace}

\def\KS      {{\ensuremath{\kaon^0_{\mathrm{S}}}}\xspace}

\def\Kstarz  {{\ensuremath{\kaon^{*0}}}\xspace}

\def\D       {{\ensuremath{\PD}}\xspace}

\def\DorDbar {\kern \thebaroffset\optbar{\kern -\thebaroffset \PD}\xspace}
\def\Dz      {{\ensuremath{\D^0}}\xspace}

\def\Dp      {{\ensuremath{\D^+}}\xspace}
\def\Dm      {{\ensuremath{\D^-}}\xspace}

\def\DpDm    {\ensuremath{\Dp {\kern -0.16em \Dm}}\xspace}

\def\Dstarp  {{\ensuremath{\D^{*+}}}\xspace}

\def\Dsm     {{\ensuremath{\D^-_\squark}}\xspace}

\def\B       {{\ensuremath{\PB}}\xspace}

\def\BorBbar {\kern \thebaroffset\optbar{\kern -\thebaroffset \PB}\xspace}

\def\Bd      {{\ensuremath{\B^0}}\xspace}

\def\BdorBdbar {\kern \thebaroffset\optbar{\kern -\thebaroffset \Bd}\xspace}

\def\Bs      {{\ensuremath{\B^0_\squark}}\xspace}

\def\BsorBsbar {\kern \thebaroffset\optbar{\kern -\thebaroffset \Bs}\xspace}

\def\jpsi     {{\ensuremath{{\PJ\mskip -3mu/\mskip -2mu\Ppsi}}}\xspace}

\def\Y#1S{\ensuremath{\PUpsilon{(#1S)}}\xspace}

\def\proton      {{\ensuremath{\Pp}}\xspace}
\def\antiproton  {{\ensuremath{\overline \proton}}\xspace}

\def\Lz          {{\ensuremath{\PLambda}}\xspace}
\def\Lbar        {{\ensuremath{\offsetoverline{\PLambda}}}\xspace}
\def\LorLbar     {\kern \thebaroffset\optbar{\kern -\thebaroffset \PLambda}\xspace}

\def\Xires       {{\ensuremath{\PXi}}\xspace}

\def\Xim         {{\ensuremath{\Xires^-}}\xspace}
\def\Xiresbar       {{\ensuremath{\offsetoverline{\Xires}}}\xspace}

\def\Lc          {{\ensuremath{\Lz^+_\cquark}}\xspace}

\def\Xicp        {{\ensuremath{\Xires^+_\cquark}}\xspace}

\def\Lb           {{\ensuremath{\Lz^0_\bquark}}\xspace}
\def\Lbbar        {{\ensuremath{\Lbar{}^0_\bquark}}\xspace}

\def\Xibz         {{\ensuremath{\Xires^0_\bquark}}\xspace}
\def\Xibm         {{\ensuremath{\Xires^-_\bquark}}\xspace}

\def\Xibbarz      {{\ensuremath{\Xiresbar{}_\bquark^0}}\xspace}

\def\BF         {{\ensuremath{\mathcal{B}}}\xspace}

\newcommand{\decay}[2]{\mbox{\ensuremath{#1\!\to #2}}\xspace} 

\def\to                 {\ensuremath{\rightarrow}\xspace}

\def\CP                {{\ensuremath{C\!P}}\xspace}

\newcommand{\ACP}{{\ensuremath{{\mathcal{A}}^{\CP}}}\xspace}

\def\AT#1     {\ensuremath{A_{\mathrm{T}}^{#1}}\xspace}           

\def\C#1      {\ensuremath{\mathcal{C}_{#1}}\xspace}                       
\def\Cp#1     {\ensuremath{\mathcal{C}_{#1}^{'}}\xspace}                    
\def\Ceff#1   {\ensuremath{\mathcal{C}_{#1}^{\mathrm{(eff)}}}\xspace}        
\def\Cpeff#1  {\ensuremath{\mathcal{C}_{#1}^{'\mathrm{(eff)}}}\xspace}       
\def\Ope#1    {\ensuremath{\mathcal{O}_{#1}}\xspace}                       
\def\Opep#1   {\ensuremath{\mathcal{O}_{#1}^{'}}\xspace}                    

\newcommand{\nospaceunit}[1]{\ensuremath{\text{#1}}}       
\newcommand{\aunit}[1]{\ensuremath{\text{\,#1}}}       

\newcommand{\tev}{\aunit{Te\kern -0.1em V}\xspace}
\newcommand{\gev}{\aunit{Ge\kern -0.1em V}\xspace}
\newcommand{\mev}{\aunit{Me\kern -0.1em V}\xspace}
\newcommand{\kev}{\aunit{ke\kern -0.1em V}\xspace}
\newcommand{\ev}{\aunit{e\kern -0.1em V}\xspace}
 
\newcommand{\mevc}{\ensuremath{\aunit{Me\kern -0.1em V\!/}c}\xspace}
\newcommand{\gevc}{\ensuremath{\aunit{Ge\kern -0.1em V\!/}c}\xspace}
\newcommand{\mevcc}{\ensuremath{\aunit{Me\kern -0.1em V\!/}c^2}\xspace}
\newcommand{\gevcc}{\ensuremath{\aunit{Ge\kern -0.1em V\!/}c^2}\xspace}

\def\mm   {\aunit{mm}\xspace}

\def\mum  {\ensuremath{\,\upmu\nospaceunit{m}}\xspace}

\def\fb   {\ensuremath{\aunit{fb}}\xspace}
\def\invfb   {\ensuremath{\fb^{-1}}\xspace}

\def\gsim{{~\raise.15em\hbox{$>$}\kern-.85em
          \lower.35em\hbox{$\sim$}~}\xspace}
\def\lsim{{~\raise.15em\hbox{$<$}\kern-.85em
          \lower.35em\hbox{$\sim$}~}\xspace}

\def\sPlot{\mbox{\em sPlot}\xspace}

\def\pt         {\ensuremath{p_{\mathrm{T}}}\xspace}

\def\ptot       {\ensuremath{p}\xspace}

\def\evtgen     {\mbox{\textsc{EvtGen}}\xspace}

\def\geant      {\mbox{\textsc{Geant4}}\xspace}

\def\photos     {\mbox{\textsc{Photos}}\xspace}

\def\pythia     {\mbox{\textsc{Pythia}}\xspace}

\def\tell1  {TELL1\xspace}
\def\ukl1   {UKL1\xspace}

\newcommand{\lhcborcid}[1]{\href{https://orcid.org/#1}{\hspace*{0.1em}\raisebox{-0.45ex}{\includegraphics[width=1em]{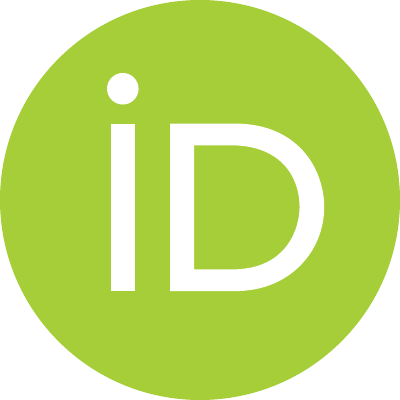}}}}


%% file: my-symbols-def.tex
\usepackage{xspace}
\usepackage{upgreek}

\def\Kstarone {{\ensuremath{K^{*}(892)^{-}}}\xspace}
\def\Kstarreszero {{\ensuremath{K^{*}_0(1430)^{-}}}\xspace}
\def\Kstarresone {{\ensuremath{K^{*}(1410)^{-}}}\xspace}
\def\Kstarrestwo {{\ensuremath{K^{*}_2(1430)^{-}}}\xspace}

\def\LbXibz {\ensuremath{\Lb(\Xibz)}\xspace}
\def\LbXibzbar {\ensuremath{\Lbbar(\Xibbarz)}\xspace}
\def\KSph {\ensuremath{\proton\KS h^-}\xspace}
\def\KSppi {\ensuremath{\proton\KS\pim}\xspace}
\def\KSpK {\ensuremath{\proton\KS\Km}\xspace}

\def\LcpiTopKS {\ensuremath{\Lc(\to\proton\KS)\pim}\xspace}
\def\LcKTopKS {\ensuremath{\Lc(\to\proton\KS)\Km}\xspace}
\def\KSpbarh {\ensuremath{\antiproton\KS h^+}\xspace}
\def\KSpbarpi {\ensuremath{\antiproton\KS\pip}\xspace}

\def\LbXibzToKSph {\decay{\LbXibz}{\KSph}}

\def\LbXibzToKSppi {\decay{\LbXibz}{\KSppi}}
\def\LbXibzToKSpK {\decay{\LbXibz}{\KSpK}}
\def\LbXibzbarToKSph {\decay{\LbXibzbar}{\KSpbarh}}

\def\LbToKSph {\decay{\Lb}{\KSph}}

\def\LbToKSppi {\decay{\Lb}{\KSppi}}
\def\LbToKSpK {\decay{\Lb}{\KSpK}}

\def\LbbarToKSppi {\decay{\Lbbar}{\KSpbarpi}}

\def\XibzToKSph {\decay{\Xibz}{\KSph}}

\def\XibzToKSppi {\decay{\Xibz}{\KSppi}}
\def\XibzToKSpK {\decay{\Xibz}{\KSpK}}

\def\LbToLchTopKS {\decay{\Lb}{\Lc(\to\proton\KS) h^{-}}}
\def\LbToLcpi {\decay{\Lb}{\Lc\pim}}
\def\LbToLcpiTopKS {\decay{\Lb}{\Lc(\to\proton\KS)\pim}}

\def\LbToLcKTopKS {\decay{\Lb}{\Lc(\to\proton\KS)\Km}}

\def\LbToppi {\decay{\Lb}{p\pim}}
\def\LbTopK {\decay{\Lb}{p\Km}}

\def\LcTopKS {\decay{\Lc}{\proton\KS}}

\def\LcTopKSres {\ensuremath{\Lc\to\proton\KS}\xspace}

\def\XicpTopKSres {\ensuremath{\Xicp\to\proton\KS}\xspace}
\def\DsToKKSres {\ensuremath{\Dsm\to\KS\Km}\xspace}
\def\DTopiKSres {\ensuremath{\Dm\to\KS\pim}\xspace}
\def\DToKKSres {\ensuremath{\Dm\to\KS\Km}\xspace}

\def\LbToKstarphToKSpiz {\decay{\Lb}{\proton\Kstarz(\to\KS\piz) h^-}}

\def\LbToSigcpiToLcpizTopKS {\decay{\Lb}{\Sigma_c^+(\to\Lc(\to\proton\KS)\piz) \pim}}

\def\XibzToKstarpKToKSpiz {\decay{\Xibz}{\proton\Kstarz(\to\KS\piz) K^-}}

\def\DstpToDzpi {\decay{\Dstarp}{\Dz(\to \Km \pip \pim \pip) \pip}}
\def\DpToKpipi {\decay{\Dp}{\Km\pip\pip}}
\def\DpToKSpi {\decay{\Dp}{\KS\pip}}

\def\XibmTojpsiXim {\decay{\Xibm}{\jpsi\Xim}}
\def\LbTojpsiLz {\decay{\Lb}{\jpsi\Lz}}
\def\DpToDzpi {\decay{\Dstarp}{\Dz\pip}}
\def\LzToppi {\decay{\Lz}{\proton\pim}}
\def\LcTopKmpip {\decay{\Lc}{\proton\Km\pip}}
\def\jpsiTomumu {\decay{\jpsi}{\mup\mun}}

\def\LbToLcmunudetail {\decay{\Lb}{\Lc(\to\proton\Km\pip)\mun\neumb}}

\def\KSTopipi {\decay{\KS}{\pip\pim}}

\def\LzToppi {\decay{\Lz}{p\pim}}

\def\BdBs {\ensuremath{\B(\Bs)}\xspace}

\def\BdsToKSKh {\decay{\Bd(\Bs)}{\KS K^+ h^-}}
\def\KSKh {\ensuremath{\KS K^+ h^-}\xspace}

\def\LbToLKK {\decay{\Lb}{\Lz\Kp\Km}}
\def\LbTopKpipi {\decay{\Lb}{\proton\Km\pip\pim}}

\def\LbTopKstar {\decay{\Lb}{\proton\Kstarone}}
\def\LbTopKstarVToKSpieg {\decay{\Lb}{\proton\Kstarone(\to\KS\pim)}}

\def\BRR {{\ensuremath{\mathcal{R}}}\xspace}

\def\Araw {\ensuremath{{\mathcal{A}}^{\mathrm{raw}}}\xspace}
\def\Ap {\ensuremath{{\mathcal{A}}^{\mathrm{P}}}\xspace}
\def\Ad {\ensuremath{{\mathcal{A}}^{\mathrm{D}}}\xspace}
\def\Apid {\ensuremath{{\mathcal{A}}^{\mathrm{PID}}}\xspace}
\def\Atrig {\ensuremath{{\mathcal{A}}^{\mathrm{trigger}}}\xspace}



%% file: title-LHCb-PAPER.tex

\begin{titlepage}
\pagenumbering{roman}

\vspace*{-1.5cm}
\centerline{\large EUROPEAN ORGANIZATION FOR NUCLEAR RESEARCH (CERN)}
\vspace*{1.5cm}
\noindent
\begin{tabular*}{\linewidth}{lc@{\extracolsep{\fill}}r@{\extracolsep{0pt}}}
\ifthenelse{\boolean{pdflatex}}
{\vspace*{-1.5cm}\mbox{\!\!\!\includegraphics[width=.14\textwidth]{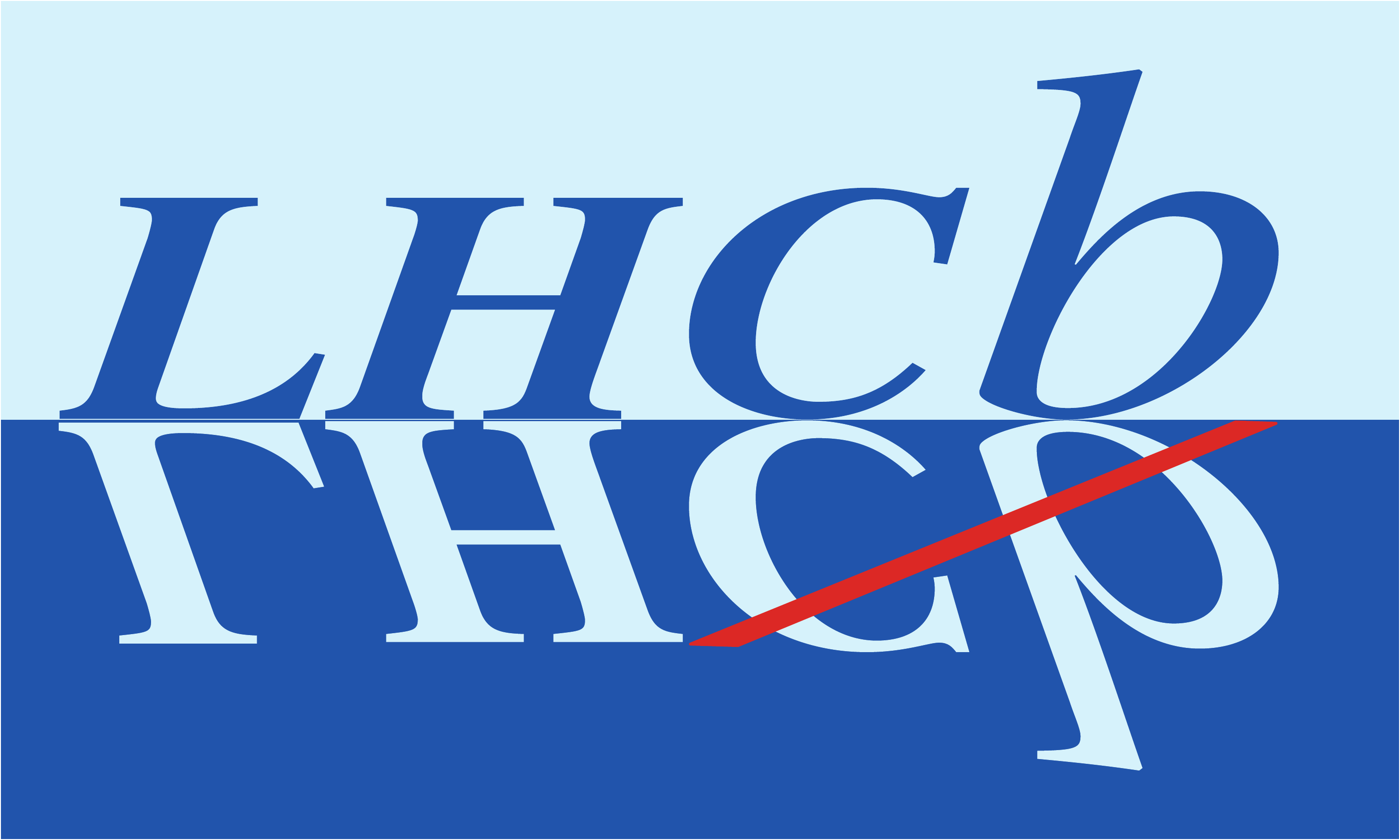}} & &}%
{\vspace*{-1.2cm}\mbox{\!\!\!\includegraphics[width=.12\textwidth]{figs/lhcb-logo.eps}} & &}%
\\
 & & CERN-EP-2025-158 \\  
 & & LHCb-PAPER-2025-016 \\  
 & & October 29, 2025 \\ 
 & & \\
\end{tabular*}

\vspace*{4.0cm}

{\normalfont\bfseries\boldmath\huge
\begin{center}
  \papertitle 
\end{center}
}

\vspace*{2.0cm}

\begin{center}
\paperauthors\footnote{Authors are listed at the end of this paper.}
\end{center}

\vspace{\fill}

\begin{abstract}
  \noindent
  A study of $\mathit{\Lambda}_b^0$ and $\mathit{\Xi}_b^0$ baryon decays to the final states $pK_{\mathrm S}^0\pi^-$ and $pK_{\mathrm S}^0K^-$ is performed using $pp$ collision data collected by the LHCb experiment, corresponding to an integrated luminosity of $9\,\mathrm{fb}^{-1}$.
  The decays \mbox{$\mathit{\Lambda}_b^0\!\to pK_{\mathrm S}^0K^-$} and \mbox{$\mathit{\Xi}_b^0\!\to pK_{\mathrm S}^0K^-$} are observed for the first time, with significances reaching eight standard deviations.
  The branching fractions and integrated $C\!P$ asymmetries are measured for the \mbox{$\mathit{\Lambda}_b^0\!\to pK_{\mathrm S}^0\pi^-$}, \mbox{$\mathit{\Lambda}_b^0\!\to pK_{\mathrm S}^0K^-$}, and \mbox{$\mathit{\Xi}_b^0\!\to pK_{\mathrm S}^0K^-$} decays. 
  For the decay \mbox{$\mathit{\Lambda}_b^0\!\to pK_{\mathrm S}^0\pi^-$}, the $C\!P$ asymmetries are measured in different regions of the Dalitz plot. 
  No evidence of $C\!P$ violation is observed.
  
\end{abstract}

\vspace*{2.0cm}

\begin{center}
  Published in
  JHEP 10 (2025) 169
\end{center}

\vspace{\fill}

{\footnotesize 
\centerline{\copyright~\papercopyright. \href{\paperlicenceurl}{\paperlicence}.}}
\vspace*{2mm}

\end{titlepage}


\newpage
\setcounter{page}{2}
\mbox{~}
%
%
%
%

%% file: body.tex
\section{Introduction}
\label{sec:introduction}

In the Standard Model (SM) of particle physics, the combined charge conjugation ($C$) and parity ($P$) symmetry is not conserved in weak interactions (\CP violation), as a consequence of an irreducible phase in the Cabibbo--Kobayashi--Maskawa (CKM) matrix~\cite{Kobayashi:1973fv}.
Although all experimental measurements of \CP violation in meson and baryon decays are compatible with the predictions of the CKM mechanism, the amount of \CP violation in the SM is insufficient to account for the observed matter-antimatter imbalance in the Universe~\cite{annurev:/content/journals/10.1146/annurev.nucl.49.1.35}.
Extensive efforts have been dedicated to searches for \CP violation in the sector of baryon decays~\cite{LHCb-PAPER-2013-061,LHCb-PAPER-2014-020,LHCb-PAPER-2016-002,LHCb-PAPER-2016-004,LHCb-PAPER-2016-030,LHCb-PAPER-2019-028,LHCb-PAPER-2016-059,LHCb-PAPER-2017-044,LHCb-PAPER-2018-001,PhysRevLett.106.181802,PhysRevLett.113.242001,LHCb-PAPER-2018-025,LHCb-PAPER-2018-044,LHCb-PAPER-2019-026,LHCb-PAPER-2020-017,LHCb-PAPER-2021-027,LHCb-PAPER-2021-030,LHCb-PAPER-2024-048}, where only direct \CP violation is expected.
Additionally, the half-integer spin of baryons introduces more partial waves in baryon decays, making them more complex compared to meson decays.
Recently, \CP violation has been observed for the first time in baryon decays using the \LbTopKpipi decay~\cite{LHCb-PAPER-2024-054},\footnote{The inclusion of charge-conjugate processes is implied throughout the article if not specified otherwise.} and evidence for \CP violation has been reported in the \LbToLKK decay~\cite{LHCb-PAPER-2024-043}.

A large \CP asymmetry is predicted in the \LbTopKstar decay, with values of $(19.7\pm1.4)\%$, $(31.1\pm2.8)\%$, and $(19.3\pm0.8)\%$ determined by the generalised factorisation approach (GFA)~\cite{PhysRevD.91.116007,PhysRevD.95.093001}, the QCD factorisation (QCDF) approach~\cite{PhysRevD.99.054020}, and the light-front quark model (LFQM) under the GFA~\cite{GENG2021136125}, respectively.
Inspired by the recent measurement of \CP asymmetries in the \LbToppi and \LbTopK decays~\cite{LHCb-PAPER-2024-048}, which are consistent with zero within $1\%$ precision, two additional predictions find a much smaller \CP asymmetry in the \LbTopKstar decay due to significant cancellation among the partial waves.
These predictions are based on the perturbative QCD (pQCD) method~\cite{han2025establishingcpviolationbbaryon} and the final-state rescattering mechanism~\cite{duan2024finalstaterescatteringmechanismbottombaryon}, and correspond to a \CP asymmetry of $(-5^{+ 10}_{- 16})\%$ and $(2\pm4)\%$, respectively.
The branching fraction of the \LbTopKstar decay is predicted to be approximately $10^{-6}$ by the GFA, the QCDF approach, and the LFQM approach, while the prediction incorporating the final-state rescattering mechanism suggests a significantly larger value of approximately $10^{-5}$.

The \LbToKSppi decay was first observed by the \lhcb experiment using proton-proton (\proton\proton) collision data corresponding to an integrated luminosity of $1\invfb$, with a \CP asymmetry of $(22\pm13\pm3)\%$~\cite{LHCb-PAPER-2013-061}, where the first and second uncertainties are statistical and systematic, respectively.
Since the \LbToKSppi decay proceeds through the quasi-two-body decay \LbTopKstarVToKSpieg, measurements of its \CP asymmetry and branching fraction provide a stringent test of the aforementioned theoretical models, and offer new insights into the \CP violation mechanism in baryon decays.

In this work, the branching fractions and the \CP-violating observables of the charmless \LbXibzToKSph decays, where $h$ denotes either a pion or a kaon, are measured using \proton\proton collision data collected by the \lhcb experiment during 2011--2012 (Run~1) and 2015--2018 (Run~2), at centre-of-mass energies of 7, 8 and 13\tev, corresponding to an integrated luminosity of $9\invfb$.
The characteristics of the \LbToKSpK, \XibzToKSppi, and \XibzToKSpK decays are similar to those of the \LbToKSppi decay, where both tree and penguin amplitudes corresponding to the $\bquark\to\uquark\uquarkbar\quark$ quark transition (with $\quark=\dquark,\squark$) contribute at leading order.
The interference between the tree and penguin amplitudes may result in \CP violation.
The \CP asymmetries in different Dalitz-plot~\cite{Dalitz:1953cp,*Fabri:1954zz} regions of the \LbToKSppi decay are also measured.
To reduce the systematic uncertainties in these measurements, the \LbToLcpiTopKS decay is used as a control channel.
The \LbToLcKTopKS decay is employed as a control sample to calibrate the contribution from misidentified background in the \LbToLcpiTopKS sample.

The paper is organised as follows.
A brief description of the \lhcb detector, trigger and simulation is presented in Sec.~\ref{sec:detector}.
The event selection is presented in Sec.~\ref{sec:selection}.
The measurements of branching fractions and \CP-violating observables are presented in Secs.~\ref{sec:bf} and~\ref{sec:acp}, respectively.
Conclusions are given in Sec.~\ref{sec:conclusion}.

\section{LHCb detector, trigger and simulation}
\label{sec:detector}

The \lhcb detector~\cite{LHCb-DP-2008-001,LHCb-DP-2014-002} is a single-arm forward
spectrometer covering the \mbox{pseudorapidity} range $2<\eta <5$,
designed for the study of particles containing \bquark or \cquark
quarks. The detector used for this work includes a high-precision tracking system
consisting of a silicon-strip vertex detector surrounding the $pp$
interaction region, a large-area silicon-strip detector located
upstream of a dipole magnet with a bending power of about
$4{\mathrm{\,T\,m}}$, and three stations of silicon-strip detectors and straw
drift tubes
placed downstream of the magnet.
The tracking system provides a measurement of the momentum, \ptot, of charged particles with
a relative uncertainty varying from 0.5\% at low momentum to 1.0\% at 200\gevc.
The minimum distance of a track to a primary $pp$ collision vertex (PV), the impact parameter (IP), 
is measured with a resolution of $(15+29/\pt)\mum$,
where \pt is the component of the momentum transverse to the beam, in\,\gevc.
Different types of charged hadrons are distinguished using information
from two ring-imaging Cherenkov detectors. 
Photons, electrons and hadrons are identified by a calorimeter system consisting of
scintillating-pad and preshower detectors, an electromagnetic
and a hadronic calorimeter. Muons are identified by a
system composed of alternating layers of iron and multiwire
proportional chambers.

The online event selection is performed by a trigger, 
which consists of a hardware stage, based on information from the calorimeter and muon
systems, followed by a software stage, which applies a full event
reconstruction.
At the hardware trigger stage, events are required to have a muon with high \pt or a
hadron, photon or electron with high transverse energy in the calorimeters. For hadrons,
the transverse energy threshold is 3.5\gev.
The software trigger requires a two-, three- or four-track
secondary vertex with a significant displacement from any primary
$pp$ interaction vertex. At least one charged particle
must satisfy $\pt > 1.6\gevc$ and be
inconsistent with originating from a PV.
A multivariate algorithm~\cite{BBDT,LHCb-PROC-2015-018} is used for the identification of secondary vertices consistent with the decay
of a \bquark hadron.

Simulation is required to model the effects of the detector acceptance, trigger, and the
imposed selection requirements.
In the simulation, $pp$ collisions are generated using
\pythia~\cite{Sjostrand:2007gs,*Sjostrand:2006za} 
with a specific \lhcb configuration~\cite{LHCb-PROC-2010-056}.
Decays of unstable particles
are described by \evtgen~\cite{Lange:2001uf}, in which final-state
radiation is generated using \photos~\cite{davidson2015photos}.
The interaction of the generated particles with the detector, and its response,
are implemented using the \geant
toolkit~\cite{Allison:2006ve, *Agostinelli:2002hh} as described in
Ref.~\cite{LHCb-PROC-2011-006}. 

\section{Event selection}
\label{sec:selection}

The candidates for \LbXibzToKSph decays are reconstructed by combining a \KS candidate with two oppositely charged tracks identified as a proton and a pion or a kaon.
The \KS candidates are selected through the \KSTopipi decay and are reconstructed in two categories.
The first category consists of \KS candidates reconstructed from two pion tracks with hits in both the vertex detector and the tracking detectors, referred to as the ``long'' \KS candidates.
The second category includes candidates reconstructed from two pion tracks with hits only in the tracking detectors, referred to as the ``downstream'' \KS candidates.
The final-state tracks (protons, pions, and kaons) are required to have a good track-fit quality and a high transverse momentum.
The \KS candidate must have a reconstructed decay vertex with a good vertex-fit quality, and a reconstructed mass within $\pm20\,(30)\mevcc$ of the known \KS mass~\cite{PDG2024}, for long (downstream) \KS meson candidates.
The \LbXibz reconstructed vertices are required to satisfy stringent vertex-fit quality criteria, while the reconstructed masses must be within $5420<m(\KSph)<6200\mevcc$.
The \KS decay vertex is selected to be downstream of the \LbXibz vertex by more than $15\mm$ along the beam.

Background from the \LzToppi decay, where the proton is misidentified as a pion, contributes to the \KS mass spectrum.
These candidates are rejected by applying a mass veto within $\pm10\mevcc$ of the known \Lz mass~\cite{PDG2024} under the corresponding mass hypothesis.
Background contributions from \BdsToKSKh decays, where a kaon is misidentified as a proton, contaminate the \LbXibz mass spectrum.
These background sources are suppressed by applying a tight particle identification (PID) requirement on the proton for \LbXibz candidates whose reconstructed mass, calculated under the kaon mass hypothesis, lies within the \BdBs mass window  $5245<m(\KSKh)<5402\mevcc$.
Decays of the \Lb or \Xibz baryons through possible charm intermediate states, such as \XicpTopKSres, \DTopiKSres, \DToKKSres and \DsToKKSres, are removed by applying mass vetoes within $\pm30\mevcc$ of the known hadron masses~\cite{PDG2024}.
For the higher-yield \LcTopKSres decay, a wider veto region of $\pm50\mevcc$ is used.

Boosted decision tree~(BDT) classifiers~\cite{Breiman,AdaBoost} are employed to further distinguish signal from background.
Separate BDT classifiers are trained for the \KSppi and \KSpK final states, for Run~1 and Run~2 periods, and for the two \KS reconstruction categories.
Simulated samples of \LbToKSph decays are used as signal training samples, while background proxies are obtained from data in the upper mass sideband region \mbox{$5900<m(\KSph)<6200\mevcc$}.
The BDT classifiers utilise the kinematic and topological variables of the \Lb and \KS hadrons and the PID variables of the proton and pion (or kaon) as inputs.
The \Lb kinematics, track multiplicity, and PID distributions in simulated samples are corrected to match those observed in data.
Two additional variables are included, both defined using tracks within a cone of radius $R=1.5$ centred on the \Lb momentum direction, excluding those used to reconstruct the \Lb candidate, with $R=\sqrt{(\Delta\eta)^2+(\Delta\phi)^2}$, where $\Delta\eta$ and $\Delta\phi$ denote the differences in pseudorapidity and azimuthal angle (in radians) between a given track and the \Lb candidate.
The first variable is the track multiplicity within the cone, and the second is the transverse momentum asymmetry between the \Lb candidate and these tracks, defined as $(\pt(\Lb)-\sum \pt({\rm track}))/(\pt(\Lb)+\sum \pt({\rm track}))$.
These two variables exploit additional information from tracks not associated with the \Lb candidate, enhancing the discrimination between signal and background.
For the \LbToKSppi decay, the selection criterion on the BDT response is optimised to suppress the combinatorial background using a figure of merit (FoM), $S/\sqrt{S+B}$, where the $S\ (B)$ denotes the estimated signal (background) yield in the \Lb signal region $5560<m(\KSph)<5680\mevcc$.
The quantities $S$ and $B$ are obtained from a fit to data with a loose BDT requirement, and scaled by the BDT selection efficiencies, determined from simulation for signal and from sideband data for background.
For the \LbToKSpK decay, which is subject to sizeable contamination from the \KSppi mode due to misidentification of the pion as a kaon, a simultaneous optimisation of the BDT response and the kaon PID requirement is performed. 
This optimisation suppresses both the combinatorial background and the misidentification contributions.
It employs the Punzi FoM~\cite{Punzi:2003bu}, $\epsilon/(a/2+\sqrt{B})$, where $\epsilon$ denotes the signal efficiency in the \Lb signal region, $a=5$ specifies the target significance of the signal component, and $B$ is the sum of combinatorial and misidentified background yields in the signal region.

The \LbToLchTopKS channels are selected using the same criteria as the \LbToKSph channels, except that the veto to the \LcTopKS decay is inverted, retaining only candidates within $\pm 30\mevcc$ of the known \Lc mass~\cite{PDG2024}.

The fraction of events with multiple \LbXibz candidates is below $1\%$.
In these events, one candidate is arbitrarily chosen.

\section{Measurement of branching fractions}
\label{sec:bf}

The ratios of branching fractions between the \LbXibzToKSph signal channels and the \LbToLcpiTopKS control channel are defined and measured according to
\begin{align}
\BRR(\LbToKSph)&\equiv\frac{\BF(\LbToKSph)}{\BF(\LbToLcpi)\BF(\LcTopKS)} \notag\\
&=\frac{N(\LbToKSph)/\epsilon(\LbToKSph)}{N(\LbToLcpiTopKS)/\epsilon(\LbToLcpiTopKS)}, \label{eq:br_Lb}\\
\BRR(\XibzToKSph)&\equiv\frac{\BF(\XibzToKSph)}{\BF(\LbToLcpi)\BF(\LcTopKS)} \notag\\
&=\frac{N(\XibzToKSph)/\epsilon(\XibzToKSph)}{N(\LbToLcpiTopKS)/\epsilon(\LbToLcpiTopKS)}\times\frac{f_{\Lb}}{f_{\Xibz}}, \label{eq:br_Xib}
\end{align}
where \BF denotes the branching fractions, $N$ denotes the yields obtained from the fit to data, $\epsilon$ denotes the corresponding total efficiencies determined using simulation, and $f_{\LbXibz}$ are the \bquark-quark fragmentation fractions~\cite{LHCb-PAPER-2018-047}.

\subsection{Mass fit}

\begin{figure}[tb]
\centering
    \includegraphics[width=0.49\linewidth]{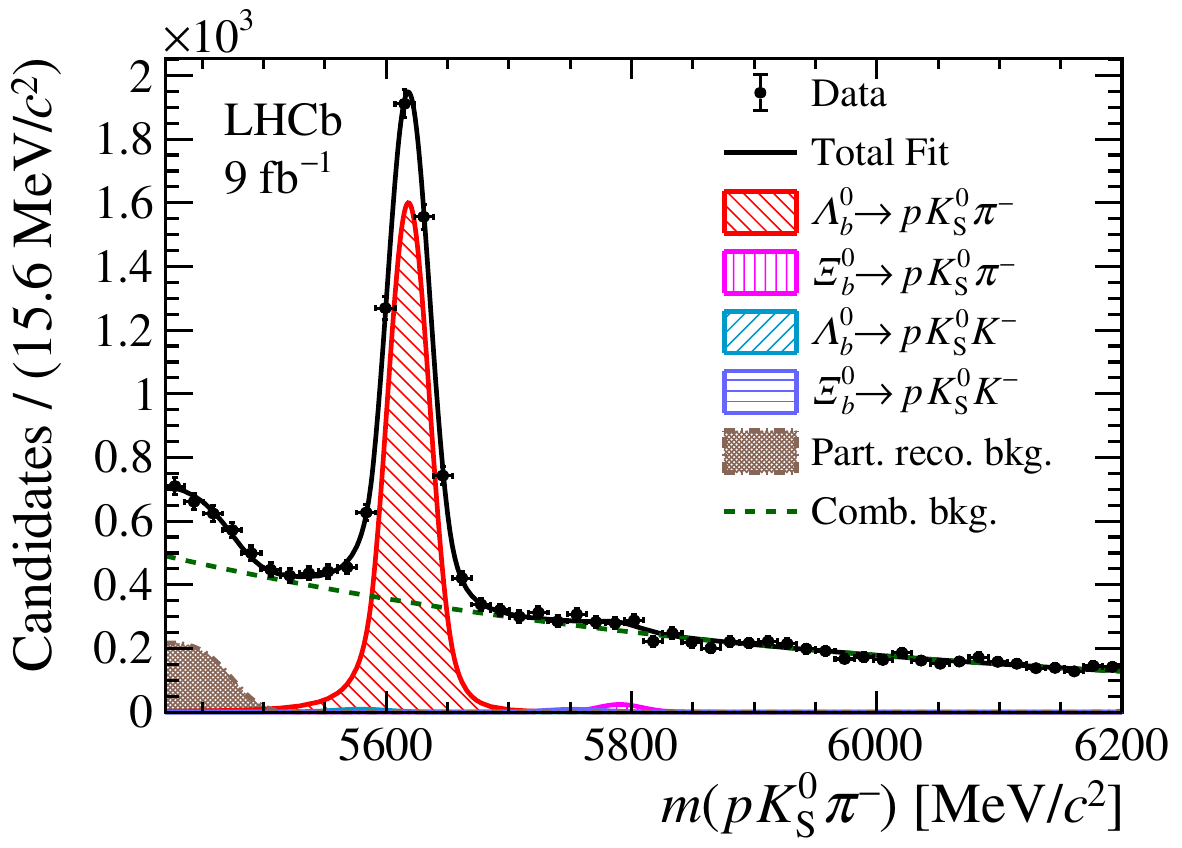}
    \includegraphics[width=0.49\linewidth]{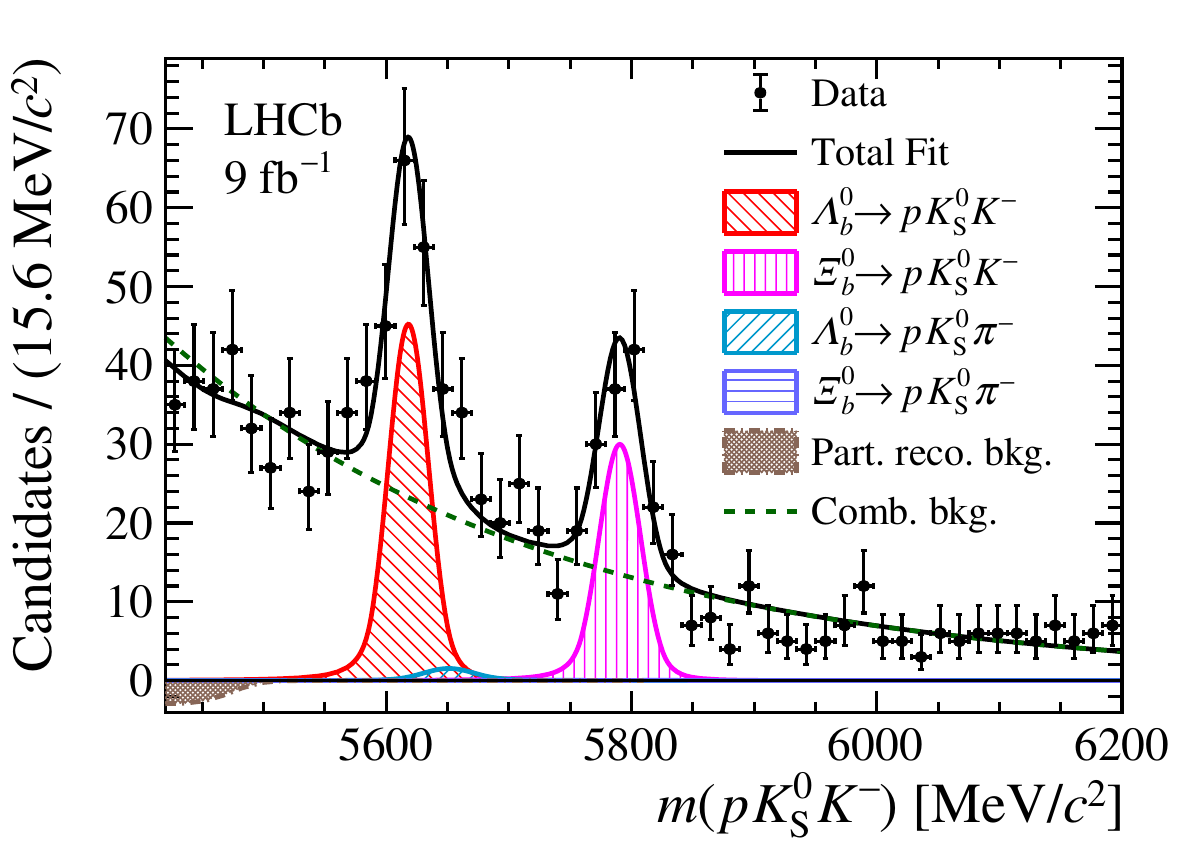}\\
    \includegraphics[width=0.49\linewidth]{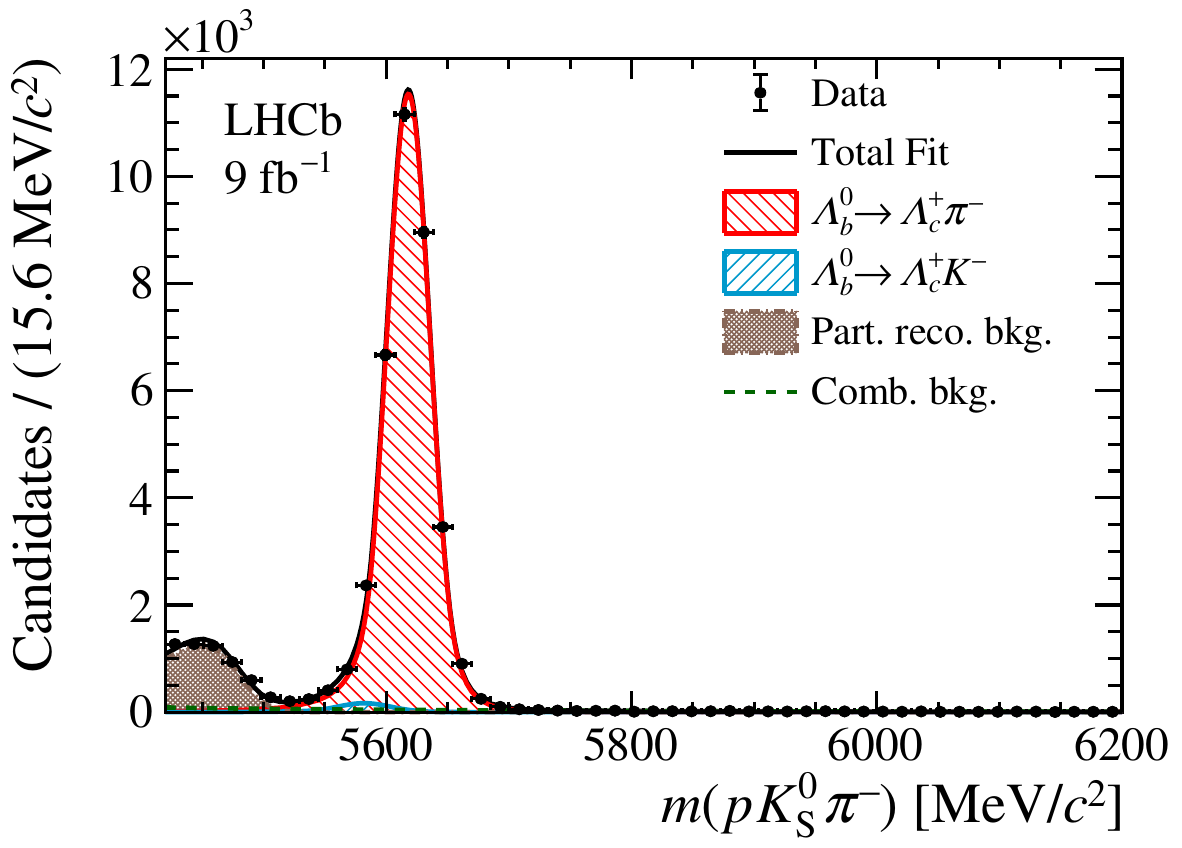}
\caption{Mass distributions for the (top left) \mbox{$\mathit{\Lambda}_b^0(\mathit{\Xi}_b^0)\!\to pK_{\mathrm S}^0\pi^-$}, (top right) \mbox{$\mathit{\Lambda}_b^0(\mathit{\Xi}_b^0)\!\to pK_{\mathrm S}^0K^-$}, and (bottom) \mbox{$\mathit{\Lambda}_b^0\!\to \mathit{\Lambda}_c^+(\to pK_{\mathrm S}^0)\pi^-$} decays, with the fit results also shown.}
\label{fig:massfit_yield}
\end{figure}

The \KSph mass distributions for the combined data of the Run~1 and Run~2 periods and the long and downstream \KS categories (four data samples) after applying all selections are presented in Fig.~\ref{fig:massfit_yield}.
A mass constraint to the known \KS mass is applied to the \KS candidates to improve the resolution of the reconstructed \KSph mass.
An extended unbinned maximum-likelihood fit is performed simultaneously to the invariant mass spectra of the four final states, \KSppi, \KSpK, \LcpiTopKS, and \LcKTopKS across the four data samples, resulting in a total of 16 samples, to extract the yields of \LbXibz signal components.
The \LbXibz signal components are modelled using a modified Gaussian function with tails on both sides of the peak (double-sided Crystal Ball~\cite{Skwarnicki:1986xj}, DSCB).
The peak position of the DSCB function, representing the mass of the \Lb or \Xibz baryon, is shared across all channels within the same data sample, with the mass difference between the \Lb and \Xibz baryons fixed to its known value~\cite{PDG2024}.
The measured \LbXibz masses are found to be consistent across the four data samples.
The width parameter of the DSCB function is parametrised as the square root of the quadratic sum of the value obtained from simulation and an additional common width term shared across all channels and data samples to account for the broader mass resolution observed in data.
The tail parameters of the DSCB function are fixed to the values obtained from fits to simulated samples.
The misidentified backgrounds from the \KSppi and \KSpK final states, due to the misidentification of a pion as a kaon or vice versa, are also modelled by DSCB functions, with their yields constrained based on the yields of the corresponding signal components and the efficiencies determined from simulation.
For the \LbToLcpiTopKS control channel, the \LbToLcKTopKS channel is used to constrain the yield of the misidentified background.
The combinatorial background is modelled using an exponential function, with a slope parameter different for each channel and data sample, except for the \LbToKSpK sample, for which the slope is shared across the four data samples due to the small sample size.
The partially reconstructed background from \Lb decays, with an additional \piz meson that is not reconstructed, is modelled by an ARGUS function~\cite{ARGUS:1990hfq} convolved with a Gaussian resolution function.
The parameter of the ARGUS function describing the mass cutoff is fixed to the known mass difference between the \Lb and \piz hadrons~\cite{PDG2024}.
The other shape parameters are free and shared across the four data samples of the same channel, except for the \LbToKSpK sample, for which the shape of the ARGUS function is fixed to that of the \LbToKSppi sample due to the small sample size.
In the fit to the \LbToKSpK sample, the yield of the partially reconstructed background is found to be negative, but statistically consistent with zero.

The fit results for the \LbXibz signal yields, reported with statistical uncertainties only, are summarised in Table~\ref{tab:signal_yield}.
For the previously unobserved \LbToKSpK, \XibzToKSppi, and \XibzToKSpK decays, the corresponding significances, evaluated using Wilks' theorem~\cite{Wilks:1938dza} and taking into account systematic uncertainties as detailed later, are also reported.
\begin{table}[tb]
\centering
\caption{Yields of the \LbXibz signal components presented with statistical uncertainties only. The corresponding significances, which include systematic uncertainties, are also provided. The significances of the previously observed \LbToKSppi and \LbToLcpiTopKS decays are not evaluated.}
\begin{tabular}{lcc}
\hline
Decay & Yield & Significance \\
\hline
\LbToKSppi & $\phantom{0}4\,740\pm90\phantom{0}$ & -- \\
\LbToKSpK & $\phantom{00\,}127\pm17\phantom{0}$ & $8.1\,\sigma$ \\
\XibzToKSppi & $\phantom{000\,}70\pm40\phantom{0}$ & $1.0\,\sigma$ \\
\XibzToKSpK & $\phantom{000\,}88\pm13\phantom{0}$ & $8.0\,\sigma$ \\
\LbToLcpiTopKS & $34\,680\pm200$ & -- \\
\hline
\end{tabular}
\label{tab:signal_yield}
\end{table}
Systematic uncertainties are incorporated by evaluating the significance under alternative fit assumptions, with the smallest value quoted as the significance including systematic effects.
Both the \LbToKSpK and the \XibzToKSpK decays are observed for the first time, with significances reaching 8 standard deviations ($\sigma$), indicating the discovery of these two decays, while no significant contribution of the \XibzToKSppi decay is observed.
Consequently, the branching fractions of the \LbToKSppi, \LbToKSpK, and \XibzToKSpK decays are measured in this work, while an upper limit is set for the \XibzToKSppi decay.

\subsection{Efficiencies}

The total efficiency, including detector acceptance, reconstruction, trigger, and offline selection efficiencies, is estimated from simulation for each channel and data sample individually.
To account for known discrepancies between simulation and data, several corrections are applied to the simulated samples.
The tracking efficiency in simulation is corrected using large calibration samples from the \jpsiTomumu decay~\cite{LHCb-DP-2013-002}.
Additionally, calibration samples from the \LzToppi, \LcTopKmpip, and \DpToDzpi decays are used to correct the mismodelling of PID information~\cite{LHCb-PUB-2016-021,LHCb-DP-2018-001} and trigger response~\cite{LHCb-DP-2012-004} in simulation.
The \Lb kinematics and track multiplicity in simulated \Lb decay samples are corrected by a weighting procedure~\cite{Rogozhnikov:2016bdp}.
The data and simulation samples of the \LbToLcpiTopKS control channel are used as input to obtain the correction weights, which are subsequently applied to all simulated \Lb decay samples.
As no control channel is available for \Xibz decays, the effect is incorporated as a systematic uncertainty for these decays.
Finally, the kinematical phase-space distribution in simulation is corrected to match that in data.
Since the \Lb baryon is unpolarised at \lhcb~\cite{LHCb-PAPER-2012-057,LHCb-PAPER-2020-005}, the phase space of the \LbToLcpiTopKS channel can be described using the \Lc helicity angle~\cite{LHCb-PAPER-2024-017}.
The simulated sample for the \LbToLcpiTopKS decay is weighted to match the measured angular distribution~\cite{LHCb-PAPER-2024-017}.
For the \LbXibzToKSph signal channels, the three-body phase space is described by the Dalitz-plot variables, $m^2(ph^-)$ and $m^2(\KS h^-)$, which are computed by constraining the reconstructed masses of the \LbXibz and \KS candidates to their known values~\cite{PDG2024}.
The background-subtracted Dalitz-plot distribution in data is obtained using the \sPlot technique~\cite{Pivk:2004ty}.
The simulated \LbXibzToKSph samples are weighted to match the dynamics of the decay.
The total efficiencies of the signal and control channels, estimated from the simulation samples after applying the same selection criteria as in data and incorporating all aforementioned corrections, are used to calculate the ratios of branching fractions.

\subsection{Systematic uncertainties}

Various sources of systematic uncertainty are taken into account for the measurement of the branching-fraction ratios.
The dominant contribution arises from the efficiency estimation, including effects from the limited size of simulation samples and corrections applied to align the PID response and phase-space distribution in simulation with those observed in data.
They are evaluated using pseudoexperiments and alternative calibration configurations, respectively.
The largest systematic uncertainty in efficiency estimation stems from the correction of the Dalitz-plot distribution in the simulation of the \LbXibzToKSph signal channels, as the distribution extracted from the background-subtracted data is subject to statistical fluctuations due to the limited signal yields.
Furthermore, for \Xibz decays, the impact of differences in \Xibz kinematics between simulation and data is considered.
The data and simulation samples are divided into regions of low and high \Xibz baryon $\pt$.
The branching-fraction ratios are measured separately in each region and found to be in agreement within the statistical uncertainty. 
The difference between the combined result of these measurements and the baseline result is assigned as a systematic uncertainty.

Three additional sources of systematic uncertainty arise from the fit to the \KSph mass spectra.
The first is due to the choice of fit models, assessed by employing alternative descriptions.
The DSCB function is replaced with a Hypatia function~\cite{Santos:2013gra}, and the exponential function with a second-order Chebyshev polynomial.
The model describing the partially reconstructed background is replaced with a shape extracted using kernel estimation~\cite{CRANMER2001198} from simulated \LbToKstarphToKSpiz and \LbToSigcpiToLcpizTopKS samples, where the \piz meson is not reconstructed, for the signal and control channels, respectively.
The second source arises from the constraint on the yield of the misidentified background, estimated by varying the efficiency ratio of misidentification and signal components within its uncertainty.
The third source is due to the potential partially reconstructed background from \Xibz decays in the \KSpK mass spectra, which is not included in the baseline fit.
The branching-fraction ratios of the \LbXibzToKSpK decays are measured by including this background, modelled using a shape extracted from a simulated \XibzToKstarpKToKSpiz sample with a missing \piz meson, and with its yield fixed to that of the \XibzToKSpK signal component.
The difference with respect to the baseline result is assigned as a systematic uncertainty.
The fit procedure is validated using pseudoexperiments, and no bias is observed.
However, the statistical uncertainties in the branching-fraction ratios of the \LbXibzToKSppi decays are found to be underestimated by $2.5\,(3.8)\%$ and are corrected accordingly.
A summary of the systematic uncertainties is provided in Table~\ref{tab:syst_ratio_bf}.
\begin{table}[tb]
\centering
\caption{Relative systematic uncertainties in the ratios of branching fractions with respect to the \LbToLcpiTopKS control channel, in percent. The total systematic uncertainties are obtained by summing the individual uncertainties in quadrature.}
\resizebox{1\columnwidth}{!}{
\begin{tabular}{l|ccccc|c}
\hline
 & Efficiency & $\Xibz$ kinematics & Model & MisID bkg. & $\Xibz$ bkg. & Total \\
\hline
$\BRR(\LbToKSppi)$ & $\phantom{0}1.4$ & -- & $0.3$ & $0.1$ & -- & $\phantom{0}1.5$ \\
$\BRR(\LbToKSpK)$ & $\phantom{0}6.4$ & -- & $3.6$ & $1.3$ & $5.7$ & $\phantom{0}9.3$ \\
$\BRR(\XibzToKSppi)\times\frac{f_{\Xibz}}{f_{\Lb}}$ & $35.8$ & $15.8$ & $1.7$ & $5.6$ & -- & $39.5$ \\
$\BRR(\XibzToKSpK)\times\frac{f_{\Xibz}}{f_{\Lb}}$ & $\phantom{0}9.1$ & $\phantom{0}5.9$ & $4.8$ & $1.0$ & $4.6$ & $12.7$ \\
\hline
\end{tabular}
}
\label{tab:syst_ratio_bf}
\end{table}

\subsection{Results}

The measured ratios of branching fractions are listed in Table~\ref{tab:ratio_bf}.
\begin{table}[tb]
\centering
\caption{Ratios of branching fractions with respect to the \LbToLcpiTopKS control channel. The first uncertainty is statistical, and the second is systematic.}
\begin{tabular}{lc}
\hline
 & Result \\
\hline
$\BRR(\LbToKSppi)$ & $0.1363\pm 0.0027\pm 0.0020$ \\
$\BRR(\LbToKSpK)$ & $0.0079\pm 0.0010\pm 0.0007$ \\
$\BRR(\XibzToKSppi)\times\frac{f_{\Xibz}}{f_{\Lb}}$ & $<0.0028\,(0.0031)\ {\rm at}\ 90\,(95)\%\ {\rm CL}$ \\
$\BRR(\XibzToKSpK)\times\frac{f_{\Xibz}}{f_{\Lb}}$ & $0.0041\pm 0.0006\pm 0.0005$ \\
\hline
\end{tabular}
\label{tab:ratio_bf}
\end{table}
Since the significance of the \XibzToKSppi decay is only $1\,\sigma$, upper limits are set at $90\%$ and $95\%$ confidence levels (CL), by integrating the profile likelihood with a uniform Bayesian prior in the region of positive branching-fraction ratio.
From the measurements of $\BF(\LbToLcpi)$ and $\BF(\LcTopKS)$~\cite{PDG2024}, and the measurement of $f_{\Xibz}/f_{\Lb}$~\cite{LHCb-PAPER-2018-047}, the absolute branching fractions of the \LbXibzToKSph decays are extracted and summarised in Table~\ref{tab:bf}.
The measured branching fraction of the \LbToKSppi decay agrees with, and supersedes, the previous measurement~\cite{LHCb-PAPER-2013-061}, with the statistical precision improved by a factor of nine.
\begin{table}[tb]
\centering
\caption{Branching fractions of the \LbXibzToKSph decays. The first uncertainty is statistical, the second is systematic, and the third arises from the uncertainty in the branching fraction of the \LbToLcpiTopKS control channel~\cite{PDG2024}. For \Xibz decays, a fourth uncertainty accounts for the \bquark-quark fragmentation fractions~\cite{LHCb-PAPER-2018-047}.}
\begin{tabular}{lc}
\hline
 & Result $[\times10^{-6}]$ \\
\hline
$\BF(\LbToKSppi)$ & $10.62\pm 0.21\pm 0.16\pm 0.98\phantom{\pm0.000}$ \\
$\BF(\LbToKSpK)$ & $\phantom{0}0.61\pm 0.08\pm 0.06\pm 0.06\phantom{\pm0.000}$ \\
$\BF(\XibzToKSppi)$ & $<2.8\,(3.2)\ {\rm at}\ 90\,(95)\%\ {\rm CL}$ \\
$\BF(\XibzToKSpK)$ & $\phantom{0}3.9\phantom{0}\pm 0.6\phantom{0}\pm 0.5\phantom{0}\pm 0.4\phantom{0}\pm 1.4\phantom{0}$ \\
\hline
\end{tabular}
\label{tab:bf}
\end{table}

\section[Measurement of CP-violating observables]{Measurement of $\boldsymbol{C\!P}$-violating observables}
\label{sec:acp}

For the \LbToKSppi, \LbToKSpK, and \XibzToKSpK signal channels, all of which have significances exceeding five standard deviations, \CP-violating observables are also measured.
The \CP asymmetry is defined as
\begin{equation}
\label{eq:def_acp}
\ACP(\LbXibzToKSph)\equiv\frac{\Gamma(\LbXibzToKSph)-\Gamma(\LbXibzbarToKSph)}{\Gamma(\LbXibzToKSph)+\Gamma(\LbXibzbarToKSph)},
\end{equation}
where $\Gamma$ represents the partial decay rate of the \LbXibzToKSph decays or their charge-conjugate processes.
The \CP asymmetry is determined by correcting the raw asymmetry for several nuisance asymmetries,
\begin{equation}
\label{eq:acp}
\ACP=\Araw-\Ap-\Ad-\Atrig-\Apid,
\end{equation}
where $\Araw=(N-\overline{N})/(N+\overline{N})$ represents the raw asymmetry in the yields between the baryon and antibaryon modes, \Ap accounts for the production asymmetry between \LbXibz and \LbXibzbar hadrons, \Ad arises from differences in detection efficiency for final-state particles, excluding the effects of trigger and PID requirements, which are given by \Atrig and \Apid, respectively.
Since the \CP asymmetry of the \LbToLcpiTopKS decay is expected to be negligible in the SM, and its characteristics are similar to those of the \LbXibzToKSph signal channels, the difference in raw asymmetries between the signal and \LbToLcpiTopKS control channels is measured to cancel most nuisance asymmetries.
The \ACP of the signal channel is therefore extracted by subtracting the residual nuisance asymmetries,
\begin{equation}
\label{eq:acp2}
\ACP=\Delta\Araw-\Delta\Ap-\Delta\Ad-\Delta\Atrig-\Delta\Apid,
\end{equation}
where $\Delta\mathcal{A}^{X}$ denotes the difference in $\mathcal{A}^{X}$ between the signal and control channels.

\subsection{Mass fit}

\begin{figure}[tb]
\centering
    \includegraphics[width=0.49\linewidth]{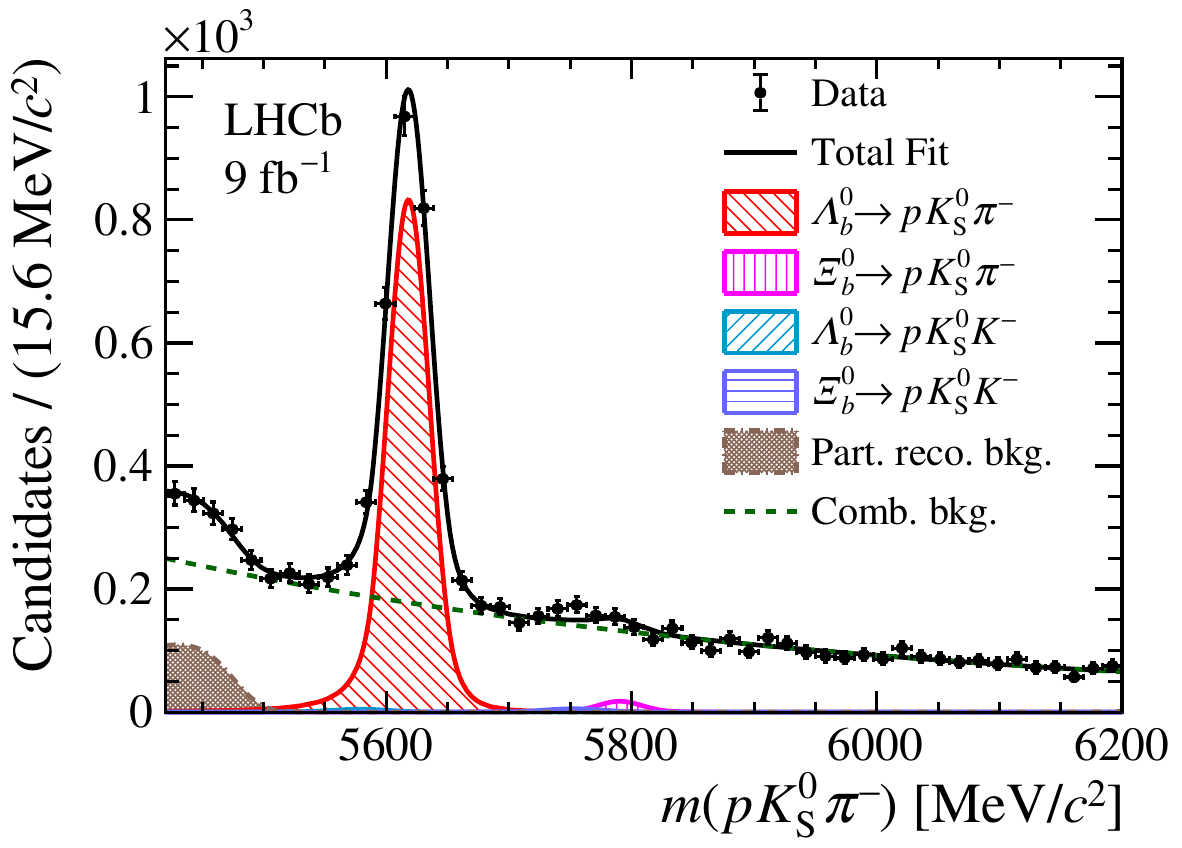}
    \includegraphics[width=0.49\linewidth]{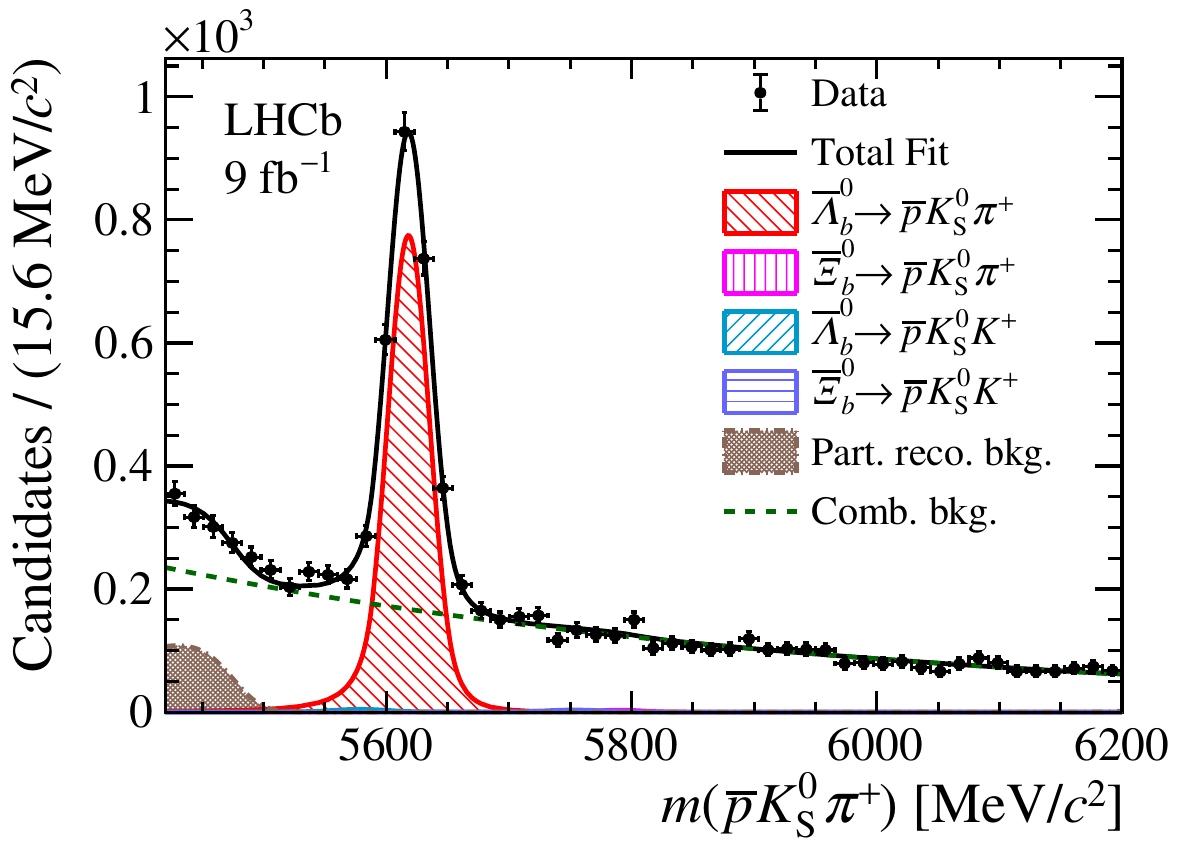}\\
    \includegraphics[width=0.49\linewidth]{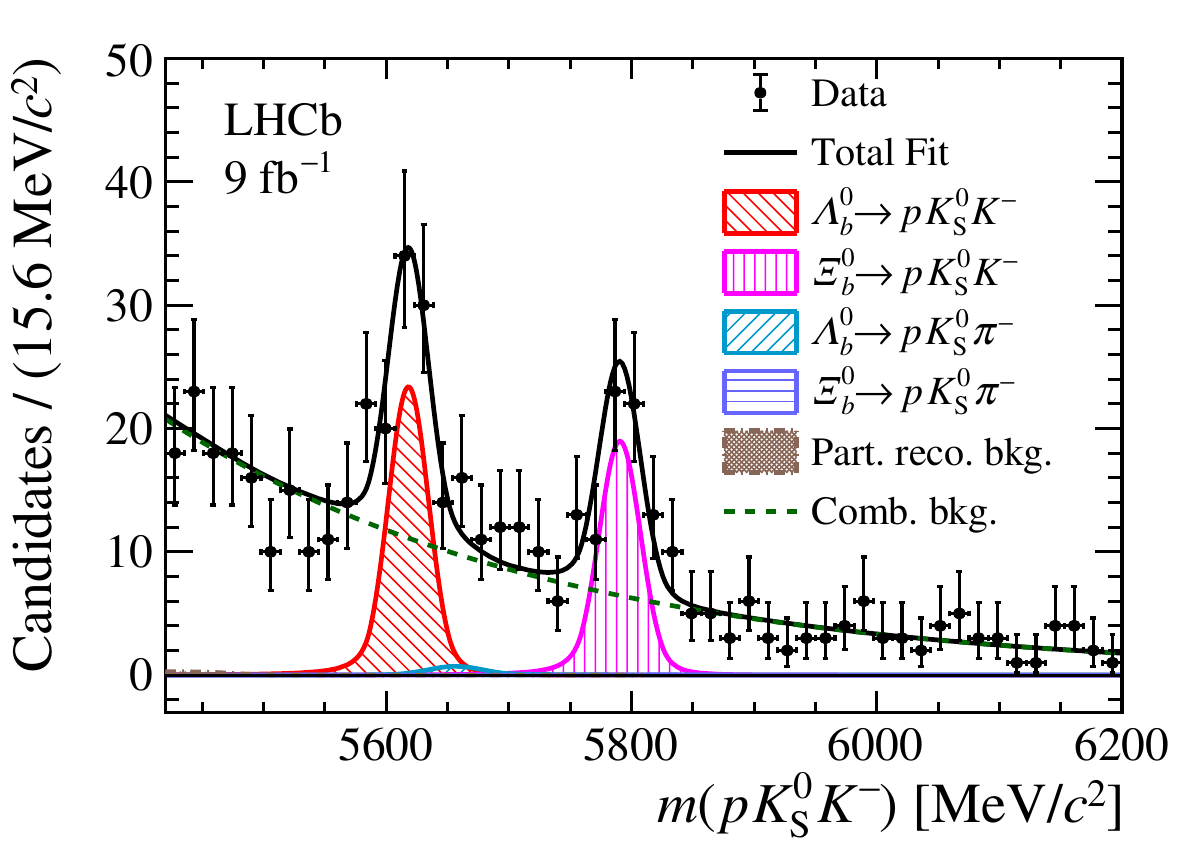}
    \includegraphics[width=0.49\linewidth]{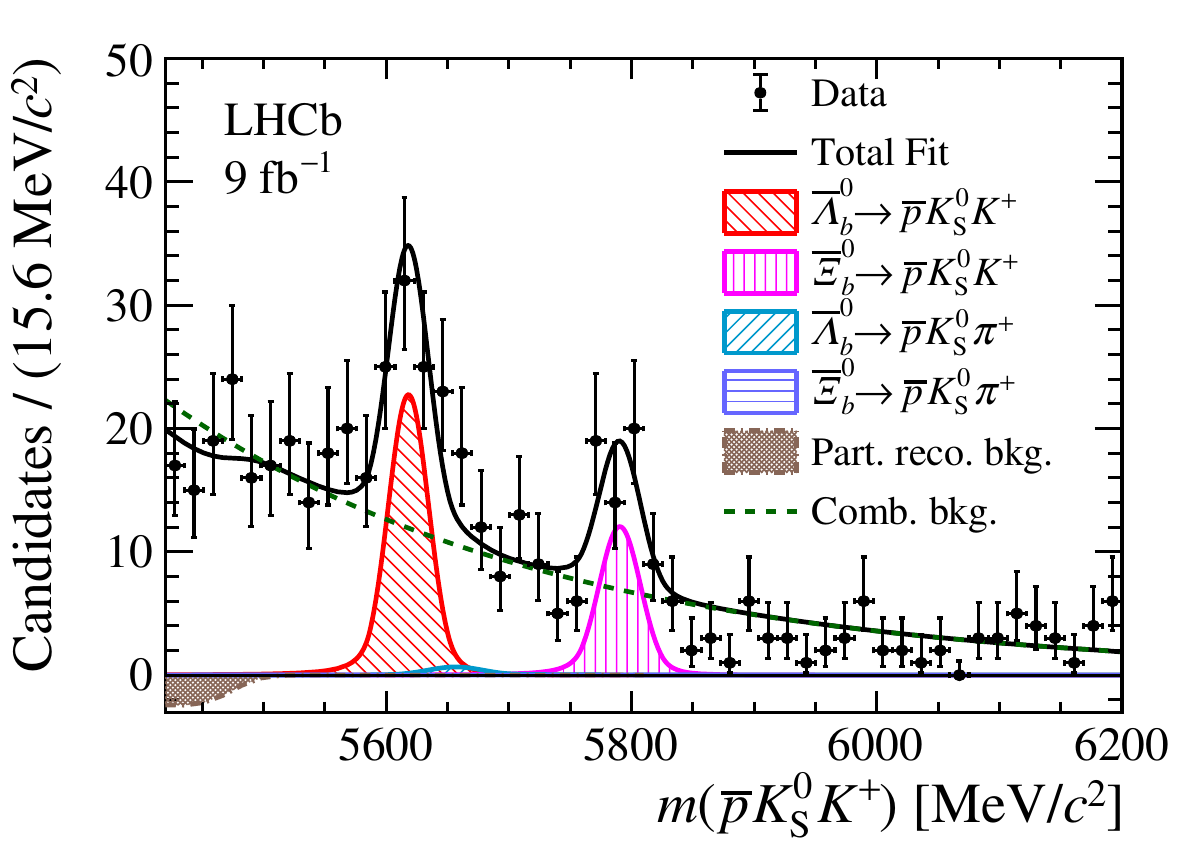}\\
    \includegraphics[width=0.49\linewidth]{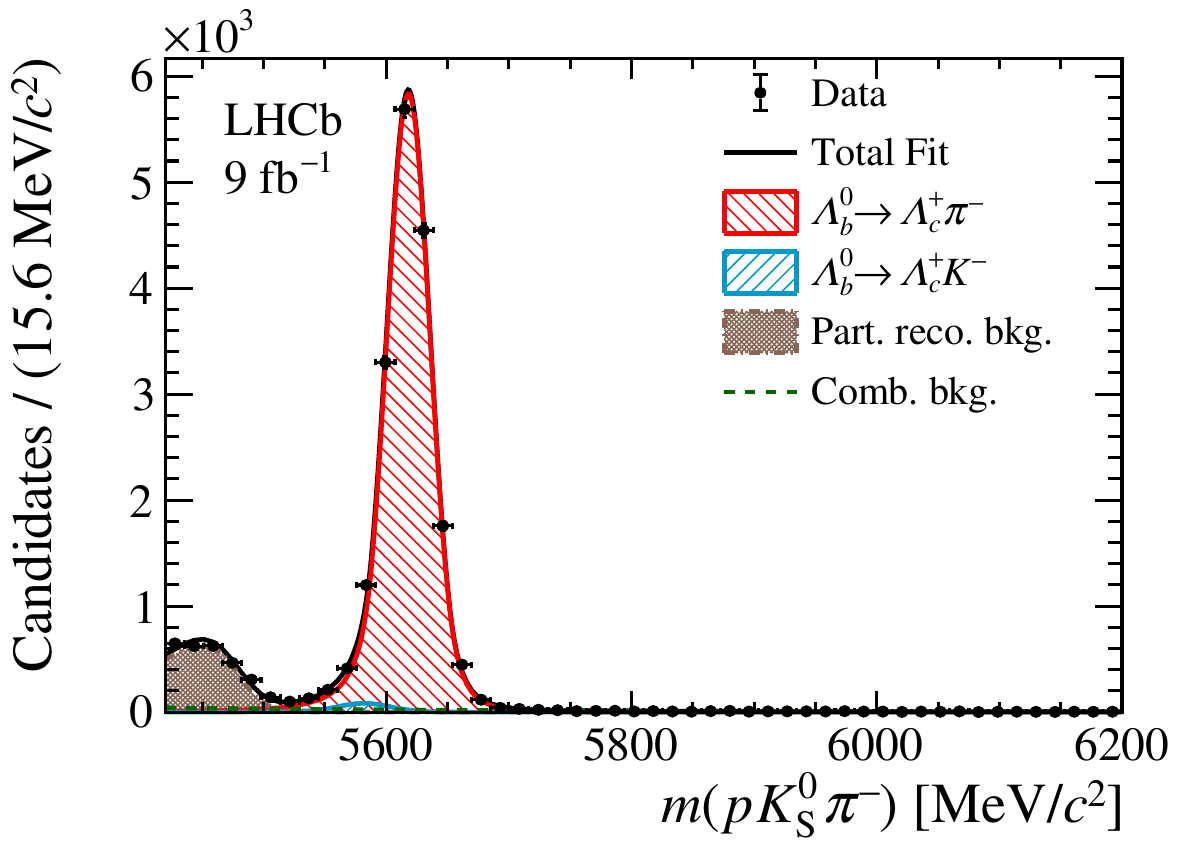}
    \includegraphics[width=0.49\linewidth]{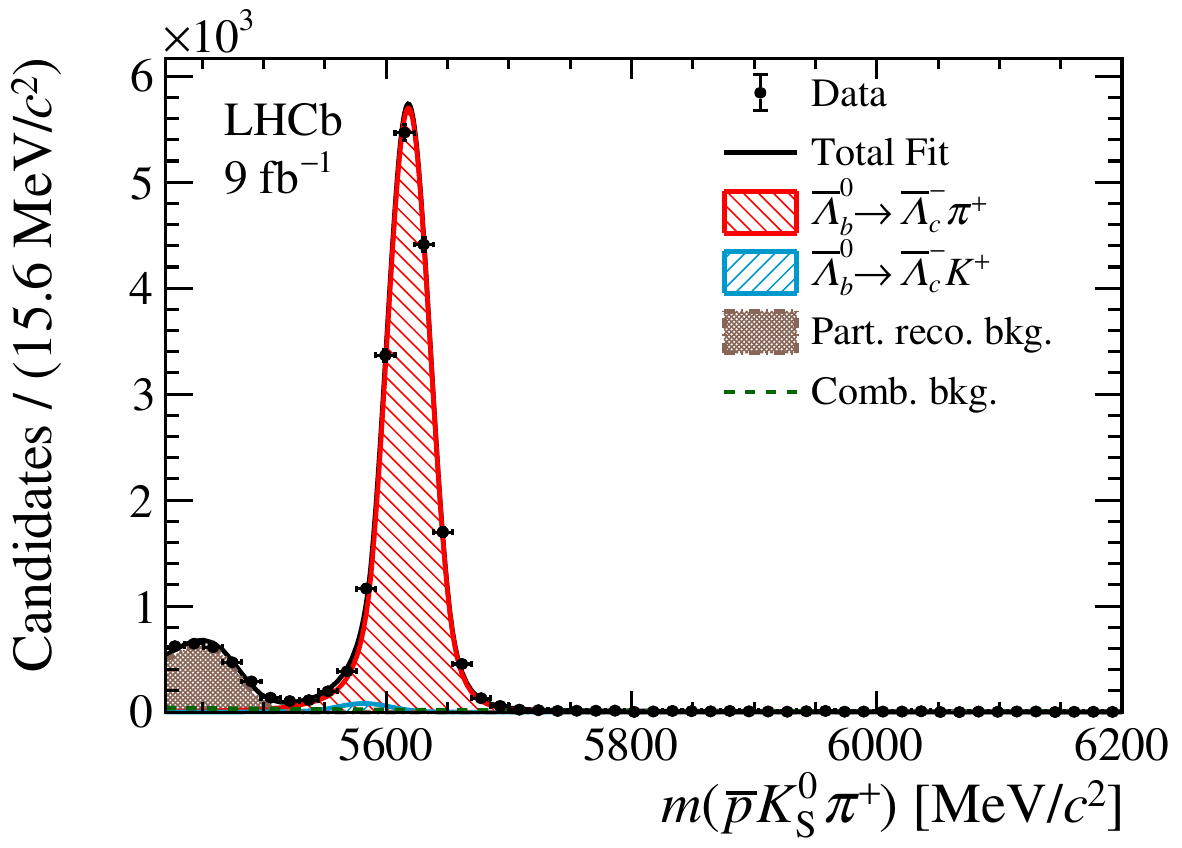}
\caption{Mass distributions for the (top) \mbox{$\mathit{\Lambda}_b^0(\mathit{\Xi}_b^0)\!\to pK_{\mathrm S}^0\pi^-$}, (middle) \mbox{$\mathit{\Lambda}_b^0(\mathit{\Xi}_b^0)\!\to pK_{\mathrm S}^0K^-$}, and (bottom) \mbox{$\mathit{\Lambda}_b^0\!\to \mathit{\Lambda}_c^+(\to pK_{\mathrm S}^0)\pi^-$} decays, shown separately for (left) baryon and (right) antibaryon samples. The fit results are also shown.}
\label{fig:massfit_daraw}
\end{figure}

The \KSph mass distributions for the combined data of the four data samples, separated into baryon and antibaryon samples after applying all selection criteria, are shown in Fig.~\ref{fig:massfit_daraw}.
An extended unbinned maximum-likelihood fit is performed simultaneously to all \KSph mass spectra of the combined data, including the final states \KSppi, \KSpK, \LcpiTopKS, \LcKTopKS, and their charge-conjugate final states, resulting in a total of eight samples.
The fit procedure follows that described in Sec.~\ref{sec:bf}, employing the same model for the corresponding components.
The yields of each component are allowed to vary independently in the fit to the baryon and antibaryon samples.
The difference in raw asymmetries between the signal and control channels is determined from this fit.
The systematic uncertainties affecting the yields are evaluated using the same methods as described in Sec.~\ref{sec:bf}.
The fit procedure is validated using pseudoexperiments, revealing a small bias in the $\Delta\Araw(\LbToKSppi)$ measurement, which is taken as a systematic uncertainty.
Additionally, the statistical uncertainties in the $\Delta\Araw(\LbToKSppi)$ and $\Delta\Araw(\XibzToKSpK)$ measurements are found to be underestimated by $2.5\%$ and $2.2\%$, respectively, and corresponding corrections are applied.
Systematic uncertainties in the fits are summarised in Table~\ref{tab:syst_acp}.
\begin{table}[tb]
\centering
\caption{Systematic uncertainties in the \ACP measurements for the \LbToKSppi, \LbToKSpK, and \XibzToKSpK decays, in percent. The total systematic uncertainties are obtained by summing the individual uncertainties in quadrature.}
\resizebox{1\columnwidth}{!}{
\begin{tabular}{cccc}
\hline
 & $\ACP(\LbToKSppi)$ & $\ACP(\LbToKSpK)$ & $\ACP(\XibzToKSpK)$\\
\hline
Model & $<0.1$ & 0.4 & \phantom{0}1.0 \\
MisID bkg. & $<0.1$ & 0.2 & \phantom{0}1.2 \\
$\Xibz$ bkg. & $\phantom{<}$-- & 7.5 & \phantom{0}5.6 \\
Fit bias & $\phantom{<}0.1$ & -- & \phantom{0}-- \\
$\Ap(\Xibz)$ & $\phantom{<}$-- & -- & \phantom{0}6.3 \\
$\Apid$ & $\phantom{<}0.6$ & 4.8 & \phantom{0}7.4 \\
Residual asym. & $\phantom{<}0.7$ & 1.6 & \phantom{0}0.6 \\
\hline
Total & $\phantom{<}0.9$ & 9.0 & 11.3 \\
\hline
\end{tabular}
}
\label{tab:syst_acp}
\end{table}

\subsection{Nuisance asymmetries}

The \Lb production asymmetry in \proton\proton collisions at centre-of-mass energies of 7 and 8\tev was measured by the \lhcb collaboration as a function of \pt and rapidity ($y$)~\cite{LHCb-PAPER-2016-062}.
The $\Ap$ values for the \LbToKSph signal and \LbToLcpiTopKS control channels are obtained by calculating the averaged \Lb production asymmetry, weighted by the two-dimensional \pt and $y$ distributions in the corresponding background-subtracted data.
Since no direct measurement of the \Xibz production asymmetry in \proton\proton collisions is available, the measurement of $\Delta\Ap$ between the \XibmTojpsiXim and the \LbTojpsiLz decays~\cite{LHCb-PAPER-2018-047} is used to estimate the systematic uncertainty in the $\ACP(\XibzToKSpK)$ measurement arising from residual production asymmetry by assuming isospin symmetry between the \Xibz and \Xibm baryons.
The detection asymmetries for charged pions and kaons in momentum bins have been measured using \DstpToDzpi, \DpToKpipi, and \DpToKSpi data~\cite{LHCb-PAPER-2016-062}.
For the detection asymmetry of protons, simulated samples of the \LbToLcmunudetail decay are used~\cite{LHCb-PAPER-2018-025}.
The $\Ad$ values for pions, kaons and protons in the signal and control channels are obtained separately by calculating the averaged detection asymmetry weighted by the momentum distribution in the background-subtracted data.
The asymmetry arising from different trigger responses to oppositely charged particles is measured using the \LbToLcmunudetail decay data, which includes pions, kaons and protons in the final state.
The PID asymmetry is considered as a systematic uncertainty, and its effect is estimated using standard \lhcb PID calibration tools~\cite{LHCb-PUB-2016-021,LHCb-DP-2018-001}.
The systematic uncertainty in estimating the residual nuisance asymmetries is evaluated using pseudoexperiments, in which the measured asymmetries in bins of the relevant variables and the corresponding distributions in the background-subtracted data are randomly varied within their uncertainties.
The resulting variation of the residual nuisance asymmetries is taken as the systematic uncertainty.
The production and detection asymmetries reported in Refs.~\cite{LHCb-PAPER-2016-062,LHCb-PAPER-2018-025} were measured using \proton\proton collision data of the Run~1 period.
These results are also applied to data of the Run~2 period in this work.
The corresponding effect is evaluated and found to be negligible for the \ACP measurement at the current precision.

\subsection[Results of integrated ACP]{Results of integrated $\boldsymbol{\ACP}$}

The \ACP values of the \LbXibzToKSph signal channels are obtained by combining the measurements of $\Delta\Araw$ and residual nuisance asymmetries, as described in Eq.~\ref{eq:acp2}, and are presented in Table~\ref{tab:acp}.
\begin{table}[tb]
\centering
\caption{\CP asymmetries \ACP for the \LbToKSppi, \LbToKSpK, and \XibzToKSpK decays. The first uncertainty is statistical, and the second is systematic.}
\begin{tabular}{lc}
\hline
 & Result [\%] \\
\hline
\ACP(\LbToKSppi) & $3.4\pm 1.9\pm 0.9$ \\
\ACP(\LbToKSpK) & $\phantom{0.}2\pm \phantom{.}13\pm \phantom{0.}9$ \\
\ACP(\XibzToKSpK) & $\phantom{.}22\pm \phantom{.}15\pm \phantom{.}11$ \\
\hline
\end{tabular}
\label{tab:acp}
\end{table}
The statistical precision of the \CP asymmetry in the \LbToKSppi decay has improved by a factor of seven in comparison with the previous \lhcb measurement~\cite{LHCb-PAPER-2013-061}.
No global \CP violation is found in any of the three signal channels.

\subsection[Results of ACP(LbToKSppi) in Dalitz-plot regions]{Results of $\boldsymbol{\ACP(\LbToKSppi)}$ in Dalitz-plot regions}

\begin{figure}[tb]
\centering
    \includegraphics[width=0.49\linewidth]{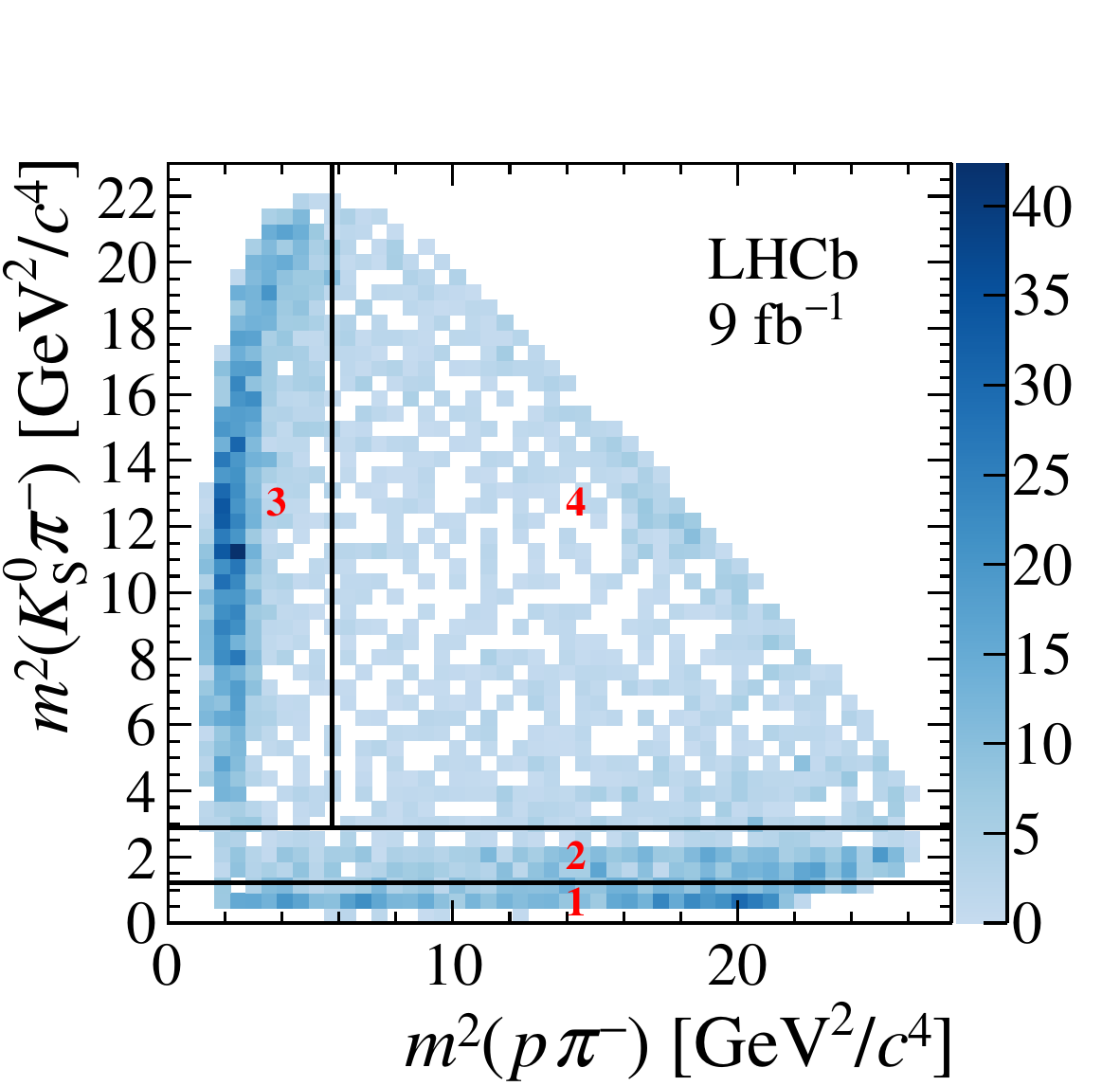}
\caption{Dalitz plot of the \mbox{$\mathit{\Lambda}_b^0\!\to pK_{\mathrm S}^0\pi^-$} decay, with background subtracted using the \mbox{$\mathit{sPlot}$} technique. The binning scheme for the  $\mathcal{A}^{C\!P}$ measurements in phase-space regions is also shown.}
\label{fig:Dalitz.bins.Lb2KSppi}
\end{figure}

\begin{figure}[tb]
\centering
    \includegraphics[width=0.49\linewidth]{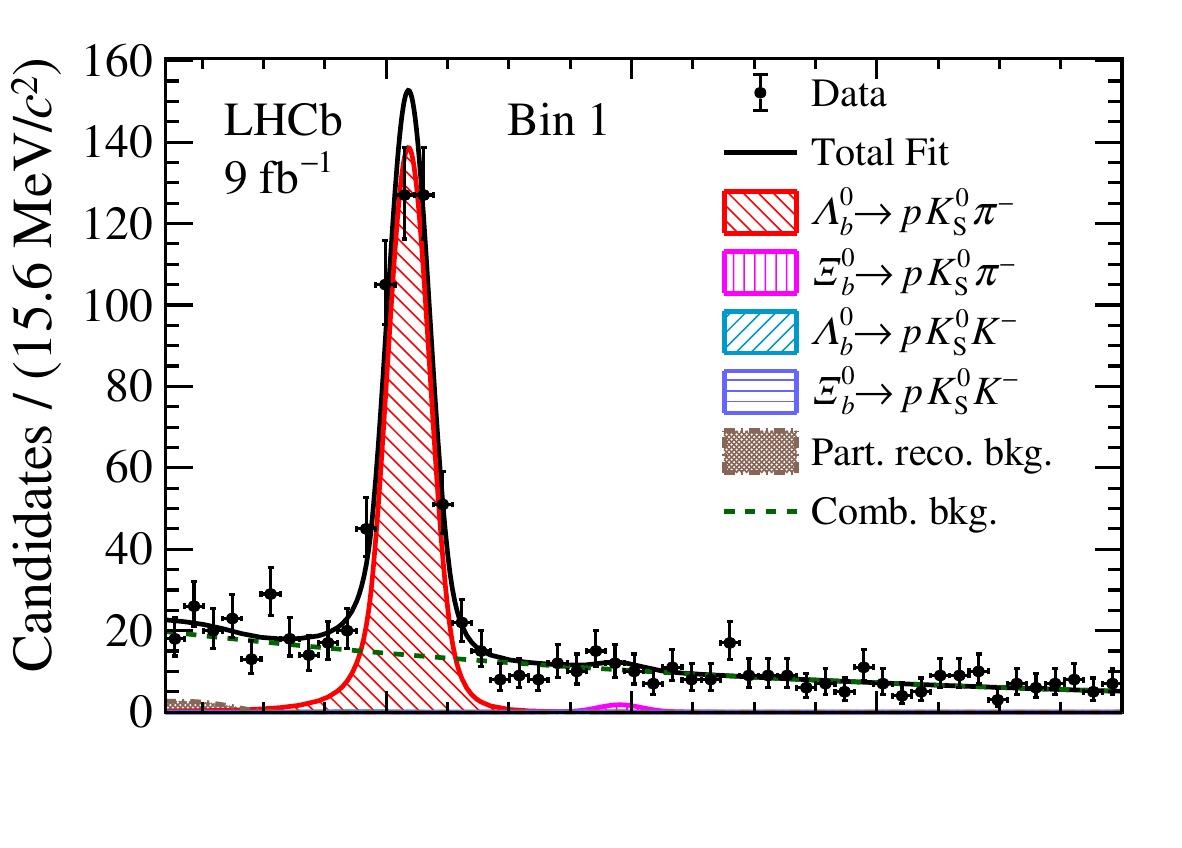}
    \includegraphics[width=0.49\linewidth]{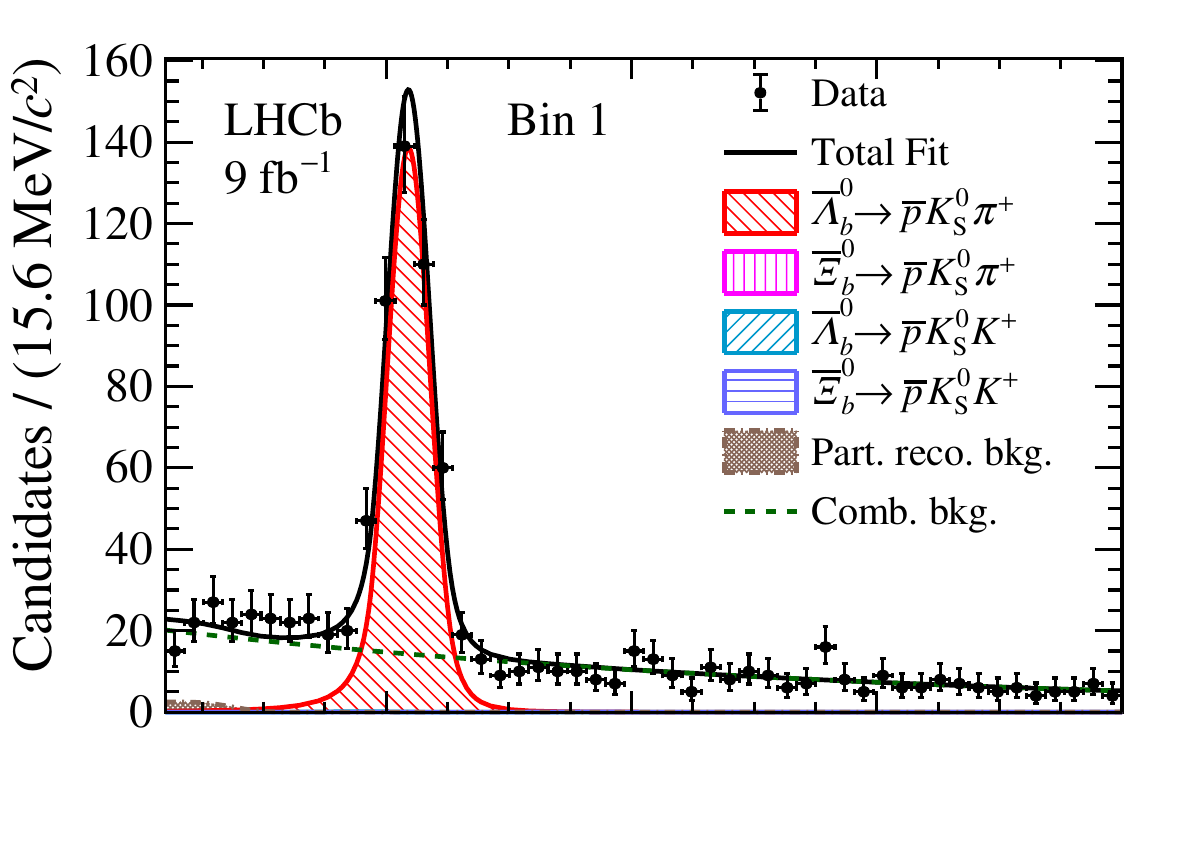}\\
    \vspace{-0.82cm}
    \includegraphics[width=0.49\linewidth]{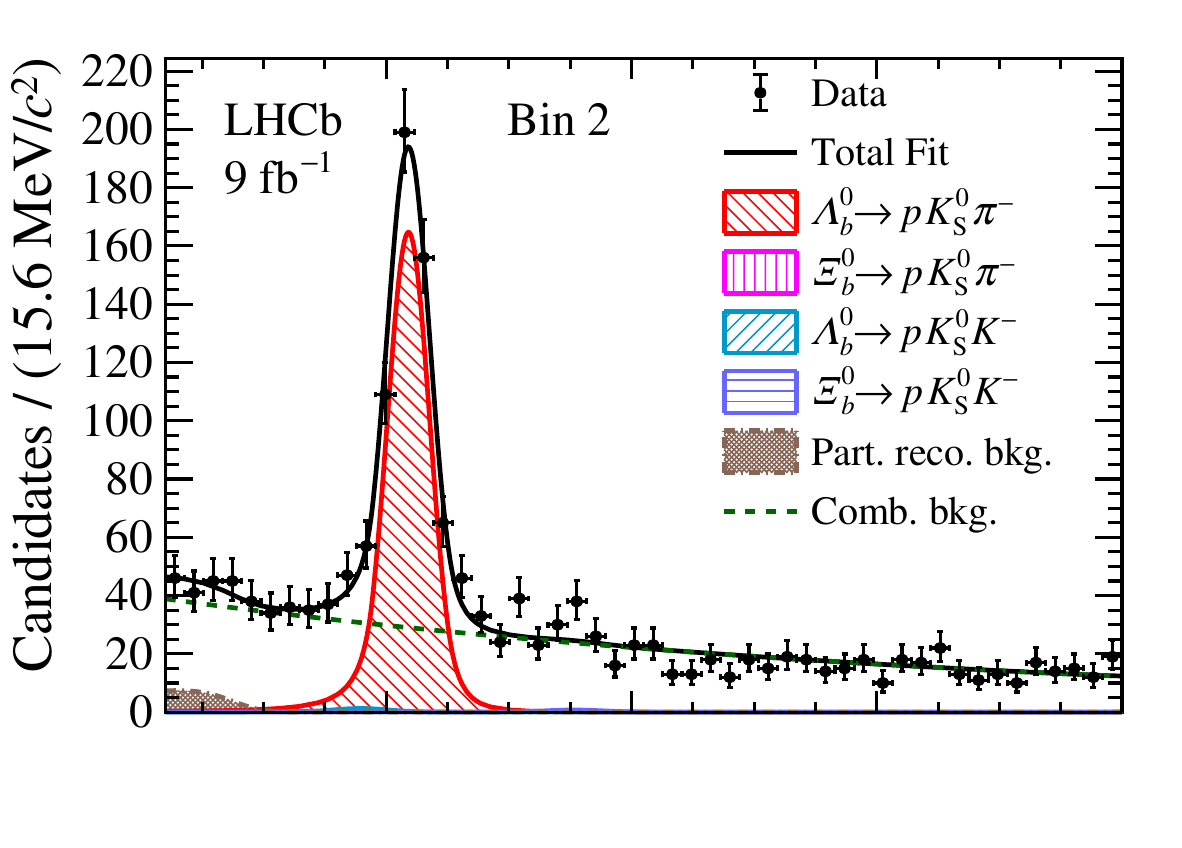}
    \includegraphics[width=0.49\linewidth]{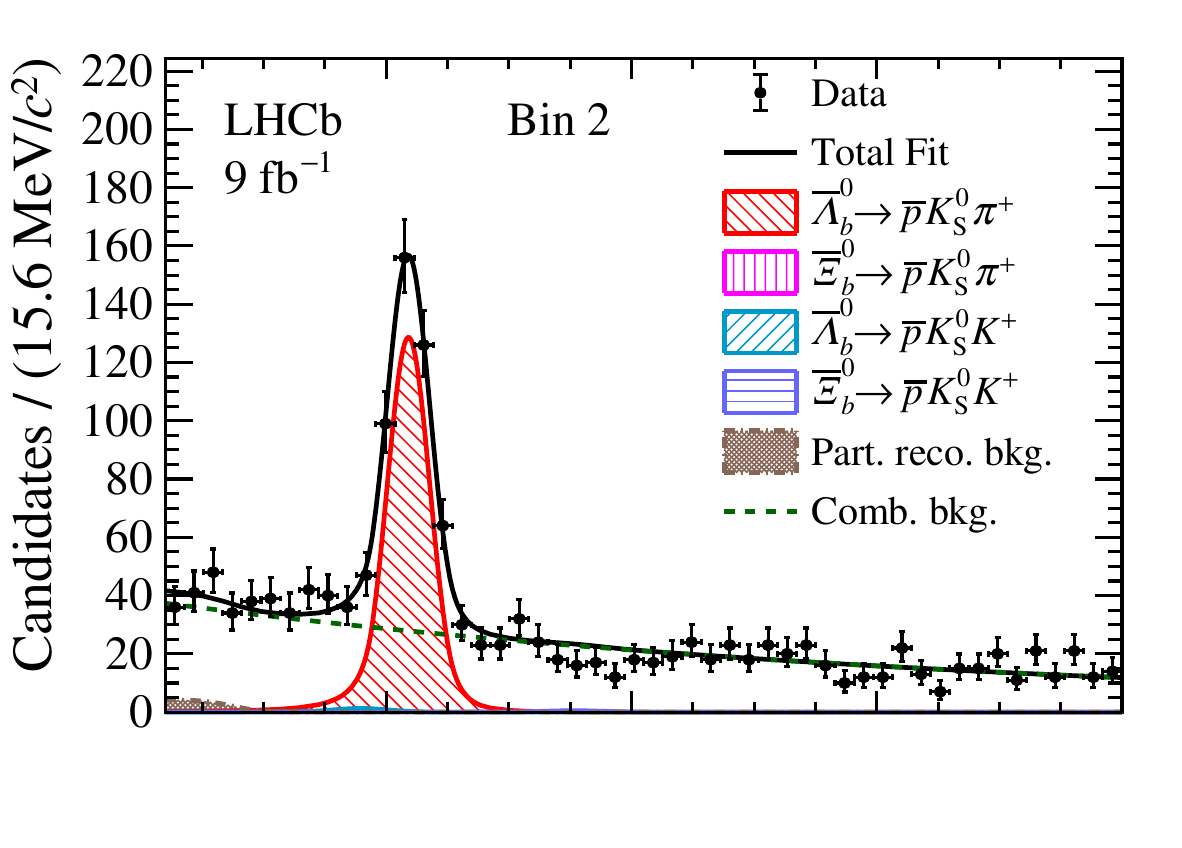}\\
    \vspace{-0.82cm}
    \includegraphics[width=0.49\linewidth]{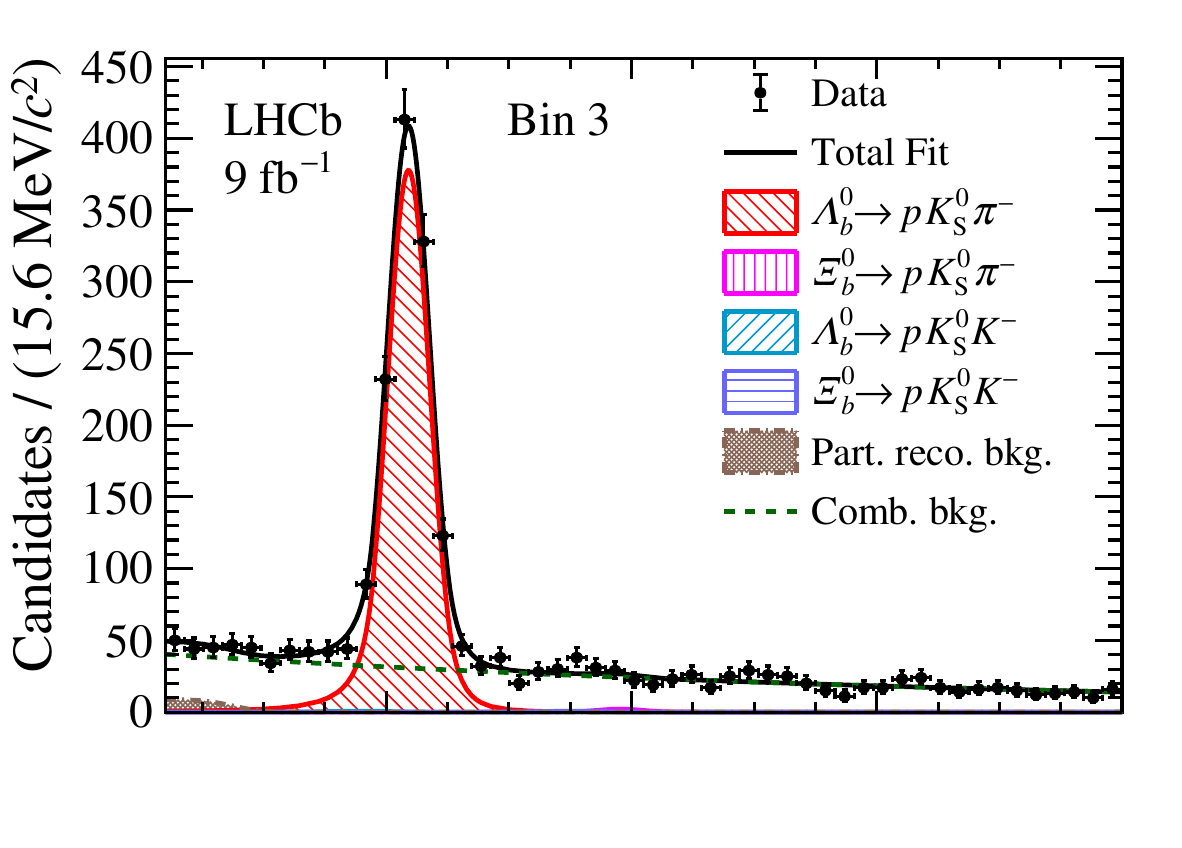}
    \includegraphics[width=0.49\linewidth]{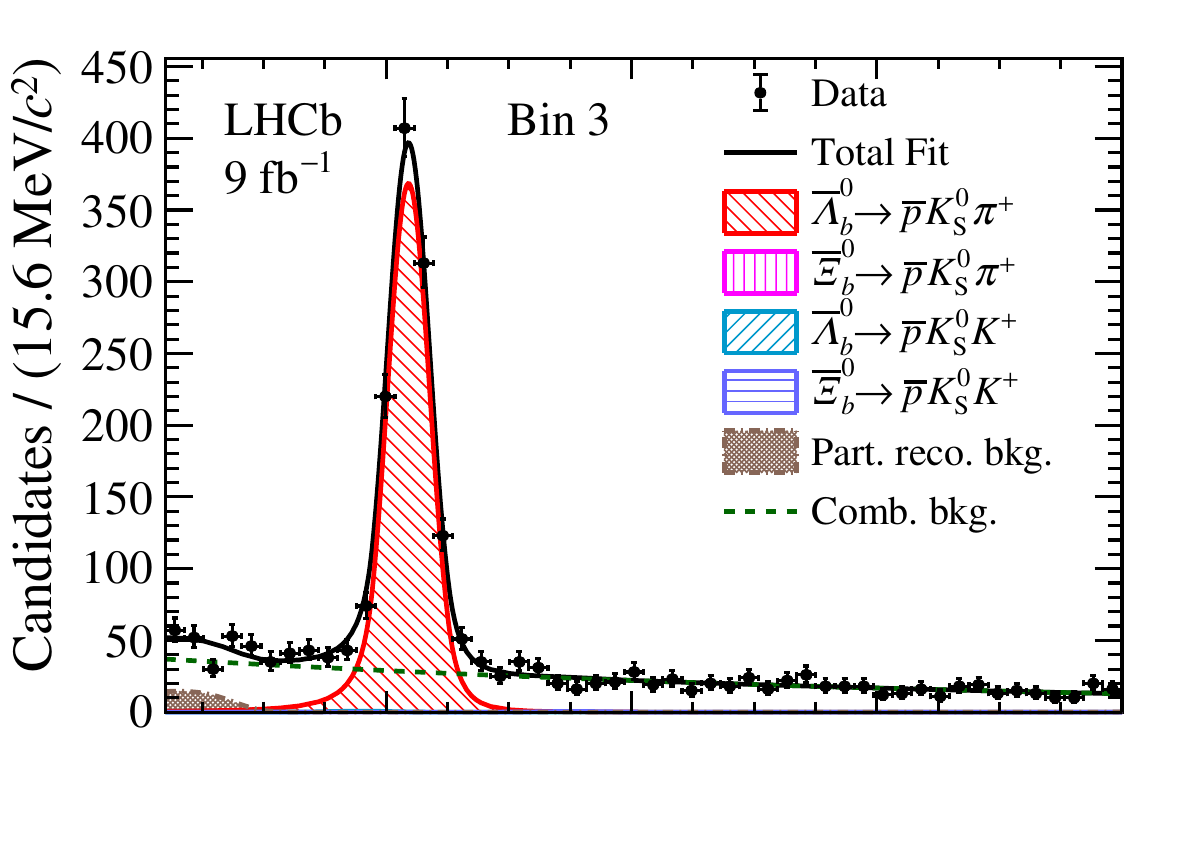}\\
    \vspace{-0.82cm}
    \includegraphics[width=0.49\linewidth]{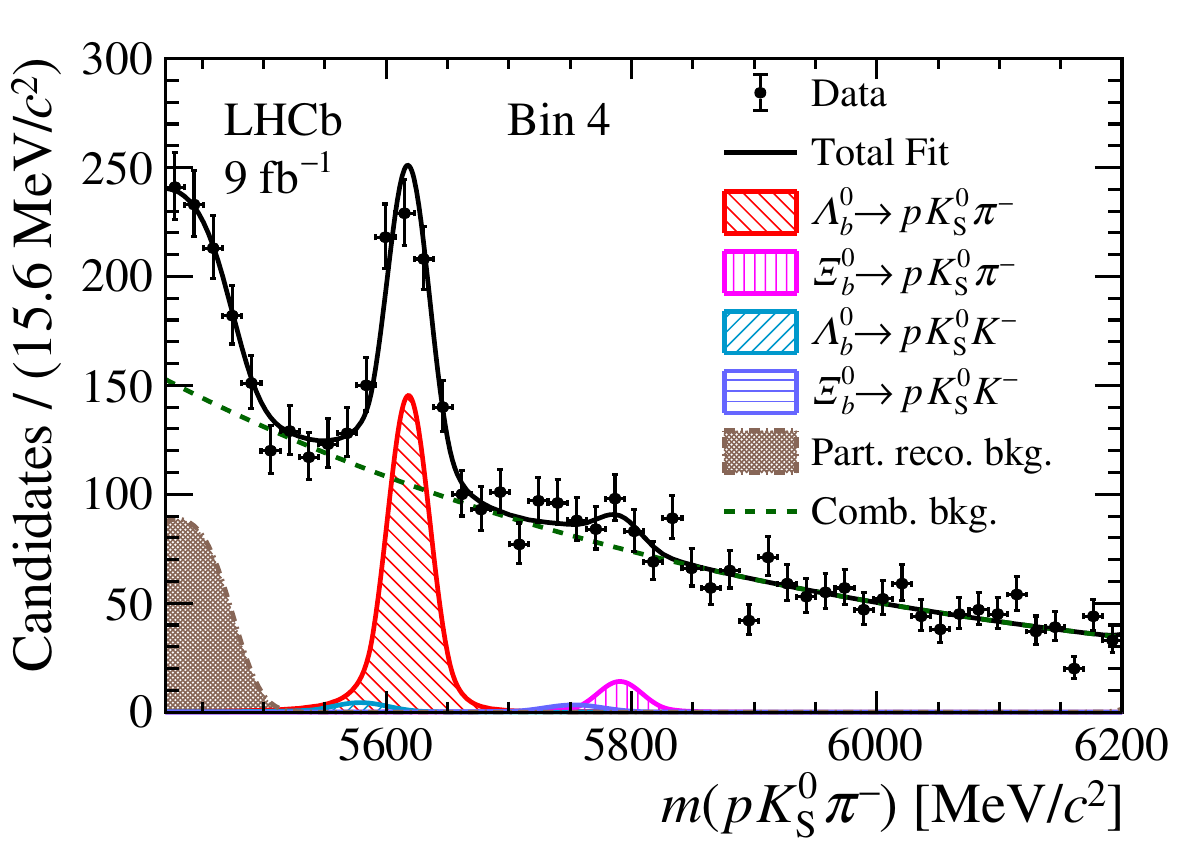}
    \includegraphics[width=0.49\linewidth]{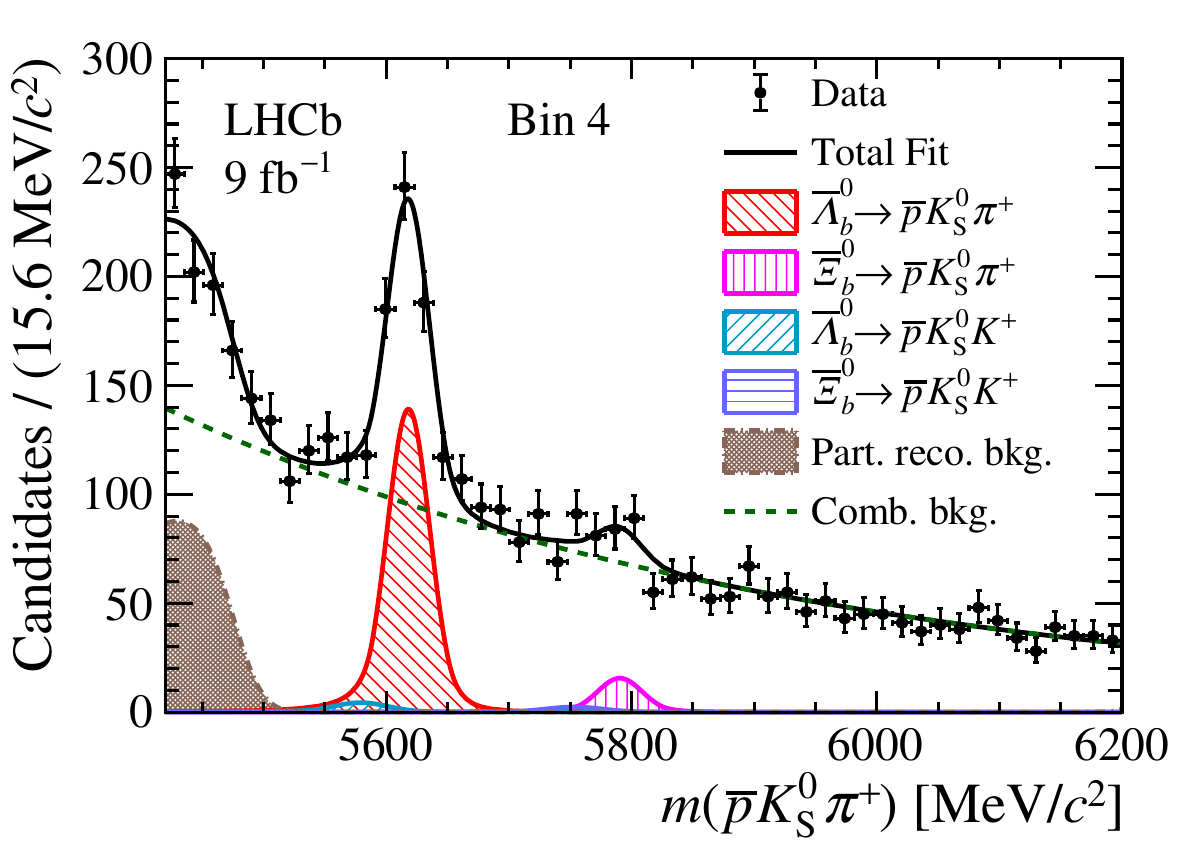}
\caption{Mass distributions for the \mbox{$\mathit{\Lambda}_b^0\!\to pK_{\mathrm S}^0\pi^-$} decay in Dalitz-plot bins 1 to 4 (from top to bottom), shown separately for (left) baryon and (right) antibaryon samples, with fit results.}
\label{fig:massfit_daraw_binned_LbToKSppi}
\end{figure}

Given the substantial signal yield of $(4.74\pm0.09)\times 10^{3}$ in the \LbToKSppi decay, the \CP asymmetry is also measured in different regions of the Dalitz plot.
Figure~\ref{fig:Dalitz.bins.Lb2KSppi} shows the Dalitz plot of the background-subtracted \LbToKSppi sample, divided into four bins according to the observed resonances: bin 1 corresponds to the \Kstarone region, bin 2 to the region of the \Kstarresone, \Kstarreszero, and \Kstarrestwo resonances, bin 3 to the $N^*$ region, and bin 4 to the nonresonant region.
The same strategy used for the integrated \ACP measurement is applied to determine the \ACP values in these bins.
For each bin, an extended unbinned maximum-likelihood fit is performed simultaneously to all \KSph mass spectra of the combined data, as in the integrated measurement.
The only difference is that the total \LbToKSppi and \LbbarToKSppi samples are replaced by their respective subsamples within the given bin, while all other samples and the fit procedure remain unchanged.
The \KSph mass distributions for the \LbToKSppi decay, divided into the four bins and further separated into baryon and antibaryon samples after applying all selection criteria, are shown in Fig.~\ref{fig:massfit_daraw_binned_LbToKSppi}.
The measured \ACP values, as well as the binning scheme, are summarised in Table~\ref{tab:acp_binned_LbToKSppi}.
\begin{table}[tb]
\centering
\caption{Measured \ACP results in different phase-space regions of the \LbToKSppi decay. The first uncertainty is statistical, and the second is systematic. The definitions for the four bins are also given. The corresponding \Lb signal yields in each bin are also reported, with their statistical uncertainties.}
\begin{tabular}{lrrrr}
\hline
 & \multicolumn{1}{c}{$m(\proton\pim)$} & \multicolumn{1}{c}{$m(\KS\pim)$} & \multicolumn{1}{c}{Yield} & \multicolumn{1}{c}{\ACP [\%]} \\
\hline
Bin 1 & \multicolumn{1}{c}{-} & $<1.1\gevcc$ & $821\pm34$ & $-0.6\pm 4.0\pm 1.9$ \\
Bin 2 & \multicolumn{1}{c}{-} & $[1.1,1.7]\gevcc$ & $870\pm40$ & $12.4\pm 4.2\pm 1.8$ \\
Bin 3 & $\leq2.4\gevcc$ & $>1.7\gevcc$ & $2200\pm50$ & $0.5\pm 2.4\pm 1.1$ \\
Bin 4 & $>2.4\gevcc$ & $>1.7\gevcc$ & $840\pm50$ & $3.3\pm 5.5\pm 2.0$ \\
\hline
\end{tabular}
\label{tab:acp_binned_LbToKSppi}
\end{table}
All the \ACP values in the four Dalitz-plot bins for the \LbToKSppi decay are consistent with zero, with the largest deviation observed in bin 2, corresponding to a significance of $2.7\,\sigma$ including systematic uncertainties.
The \ACP value in the \Kstarone dominated region is $(-0.6\pm4.0\pm1.9)\%$, where the first and second uncertainties are statistical and systematic, significantly smaller than the predicted value of $20\%$~\cite{PhysRevD.91.116007,PhysRevD.95.093001,PhysRevD.99.054020,GENG2021136125}.
A possible explanation for this result is the cancellation of \CP asymmetries among the partial waves in the \LbTopKstar decay~\cite{han2025establishingcpviolationbbaryon,duan2024finalstaterescatteringmechanismbottombaryon}.

\section{Conclusion}
\label{sec:conclusion}

Using \proton\proton collision data collected by the \lhcb experiment corresponding to an integrated luminosity of $9\invfb$, the branching fractions for the \LbToKSppi, \LbToKSpK, and \XibzToKSpK decays are measured.
The \LbToKSpK and \XibzToKSpK decays are observed for the first time, with significances reaching $8\,\sigma$.
For the \XibzToKSppi decay, since the significance is less than $3\,\sigma$, upper limits are set.
Integrated \CP asymmetries are measured for all decays except for the \XibzToKSppi decay given the limited significance. 
Additionally, \CP violation is studied in different Dalitz-plot regions of the \LbToKSppi decay.
No \CP violation is found in any of the signal decays with the current precision.
The vanishing \CP asymmetry observed for the \LbToKSppi decay in the \Kstarone mass region, significantly smaller than the approximately $20\%$ effect predicted by the GFA~\cite{PhysRevD.91.116007,PhysRevD.95.093001}, the QCDF approach~\cite{PhysRevD.99.054020}, and the LFQM approach~\cite{GENG2021136125}, is in agreement with the hypothesis of a cancellation mechanism among the contributing partial waves, as proposed in Refs.~\cite{han2025establishingcpviolationbbaryon,duan2024finalstaterescatteringmechanismbottombaryon}.
Further studies are needed to understand the mechanism of \CP violation in baryon decays.

%% file: acknowledgements.tex
\section*{Acknowledgements}
%
%
\noindent We express our gratitude to our colleagues in the CERN
accelerator departments for the excellent performance of the LHC. We
thank the technical and administrative staff at the LHCb
institutes.
We acknowledge support from CERN and from the national agencies:
ARC (Australia);
CAPES, CNPq, FAPERJ and FINEP (Brazil); 
MOST and NSFC (China); 
CNRS/IN2P3 (France); 
BMBF, DFG and MPG (Germany); 
INFN (Italy); 
NWO (Netherlands); 
MNiSW and NCN (Poland); 
MCID/IFA (Romania); 
MICIU and AEI (Spain);
SNSF and SER (Switzerland); 
NASU (Ukraine); 
STFC (United Kingdom); 
DOE NP and NSF (USA).
We acknowledge the computing resources that are provided by ARDC (Australia), 
CBPF (Brazil),
CERN, 
IHEP and LZU (China),
IN2P3 (France), 
KIT and DESY (Germany), 
INFN (Italy), 
SURF (Netherlands),
Polish WLCG (Poland),
IFIN-HH (Romania), 
PIC (Spain), CSCS (Switzerland), 
and GridPP (United Kingdom).
We are indebted to the communities behind the multiple open-source
software packages on which we depend.
Individual groups or members have received support from
Key Research Program of Frontier Sciences of CAS, CAS PIFI, CAS CCEPP, 
Fundamental Research Funds for the Central Universities,  and Sci.\ \& Tech.\ Program of Guangzhou (China);
Minciencias (Colombia);
EPLANET, Marie Sk\l{}odowska-Curie Actions, ERC and NextGenerationEU (European Union);
A*MIDEX, ANR, IPhU and Labex P2IO, and R\'{e}gion Auvergne-Rh\^{o}ne-Alpes (France);
Alexander-von-Humboldt Foundation (Germany);
ICSC (Italy); 
Severo Ochoa and Mar\'ia de Maeztu Units of Excellence, GVA, XuntaGal, GENCAT, InTalent-Inditex and Prog.~Atracci\'on Talento CM (Spain);
SRC (Sweden);
the Leverhulme Trust, the Royal Society and UKRI (United Kingdom).

%% file: Authorship_LHCb-PAPER-2025-016.tex
\centerline
{\large\bf LHCb collaboration}
\begin
{flushleft}
\small
R.~Aaij$^{38}$\lhcborcid{0000-0003-0533-1952},
A.S.W.~Abdelmotteleb$^{57}$\lhcborcid{0000-0001-7905-0542},
C.~Abellan~Beteta$^{51}$\lhcborcid{0009-0009-0869-6798},
F.~Abudin{\'e}n$^{57}$\lhcborcid{0000-0002-6737-3528},
T.~Ackernley$^{61}$\lhcborcid{0000-0002-5951-3498},
A. A. ~Adefisoye$^{69}$\lhcborcid{0000-0003-2448-1550},
B.~Adeva$^{47}$\lhcborcid{0000-0001-9756-3712},
M.~Adinolfi$^{55}$\lhcborcid{0000-0002-1326-1264},
P.~Adlarson$^{84}$\lhcborcid{0000-0001-6280-3851},
C.~Agapopoulou$^{14}$\lhcborcid{0000-0002-2368-0147}
% ...

{\footnotesize \it

% ...
$^{14}$Universit{\'e} Paris-Saclay, CNRS/IN2P3, IJCLab, Orsay, France\\
% ...
$^{38}$Nikhef National Institute for Subatomic Physics, Amsterdam, Netherlands\\
% ...
$^{47}$Instituto Galego de F{\'\i}sica de Altas Enerx{\'\i}as (IGFAE), Universidade de Santiago de Compostela, Santiago de Compostela, Spain\\
% ...
$^{51}$Physik-Institut, Universit{\"a}t Z{\"u}rich, Z{\"u}rich, Switzerland\\
% ...
$^{55}$H.H. Wills Physics Laboratory, University of Bristol, Bristol, United Kingdom\\
% ...
$^{57}$Department of Physics, University of Warwick, Coventry, United Kingdom\\
% ...
$^{60}$School of Physics and Astronomy, University of Glasgow, Glasgow, United Kingdom\\
$^{61}$Oliver Lodge Laboratory, University of Liverpool, Liverpool, United Kingdom\\
% ...
$^{69}$Syracuse University, Syracuse, NY, United States\\
% ...
$^{84}$Department of Physics and Astronomy, Uppsala University, Uppsala, Sweden, associated to $^{60}$\\
% ...
}
\end{flushleft}

%% file: main.bbl
\ifx\mcitethebibliography\mciteundefinedmacro
\PackageError{LHCb.bst}{mciteplus.sty has not been loaded}
{This bibstyle requires the use of the mciteplus package.}\fi
\providecommand{\href}[2]{#2}
\begin{mcitethebibliography}{10}
\mciteSetBstSublistMode{n}
\mciteSetBstMaxWidthForm{subitem}{\alph{mcitesubitemcount})}
\mciteSetBstSublistLabelBeginEnd{\mcitemaxwidthsubitemform\space}
{\relax}{\relax}

\bibitem{Kobayashi:1973fv}
M.~Kobayashi and T.~Maskawa, \ifthenelse{\boolean{articletitles}}{\emph{{\CP-violation in the renormalizable theory of weak interaction}}, }{}\href{https://doi.org/10.1143/PTP.49.652}{Prog.\ Theor.\ Phys.\  \textbf{49} (1973) 652}\relax
\mciteBstWouldAddEndPuncttrue
\mciteSetBstMidEndSepPunct{\mcitedefaultmidpunct}
{\mcitedefaultendpunct}{\mcitedefaultseppunct}\relax
\EndOfBibitem
\bibitem{annurev:/content/journals/10.1146/annurev.nucl.49.1.35}
A.~Riotto and M.~Trodden, \ifthenelse{\boolean{articletitles}}{\emph{Recent progress in baryogenesis}, }{}\href{https://doi.org/https://doi.org/10.1146/annurev.nucl.49.1.35}{Ann.\ Rev.\ Nucl.\ Part.\ Sci.\  \textbf{49} (1999) 35}, \href{http://arxiv.org/abs/hep-ph/9901362}{{\normalfont\ttfamily arXiv:hep-ph/9901362}}\relax
\mciteBstWouldAddEndPuncttrue
\mciteSetBstMidEndSepPunct{\mcitedefaultmidpunct}
{\mcitedefaultendpunct}{\mcitedefaultseppunct}\relax
\EndOfBibitem
\bibitem{LHCb-PAPER-2013-061}
LHCb collaboration, R.~Aaij {\em et~al.}, \ifthenelse{\boolean{articletitles}}{\emph{{Searches for \Lb and \Xibz decays to $\KS\proton\pim$ and $\KS\proton\Km$ final states with first observation of the \mbox{\decay{\Lb}{\KS\proton\pim}} decay}}, }{}\href{https://doi.org/10.1007/JHEP04(2014)087}{JHEP \textbf{04} (2014) 087}, \href{http://arxiv.org/abs/1402.0770}{{\normalfont\ttfamily arXiv:1402.0770}}\relax
\mciteBstWouldAddEndPuncttrue
\mciteSetBstMidEndSepPunct{\mcitedefaultmidpunct}
{\mcitedefaultendpunct}{\mcitedefaultseppunct}\relax
\EndOfBibitem
\bibitem{LHCb-PAPER-2014-020}
LHCb collaboration, R.~Aaij {\em et~al.}, \ifthenelse{\boolean{articletitles}}{\emph{{Observation of the \mbox{\decay{\Lb}{\jpsi\proton\pim}} decay}}, }{}\href{https://doi.org/10.1007/JHEP07(2014)103}{JHEP \textbf{07} (2014) 103}, \href{http://arxiv.org/abs/1406.0755}{{\normalfont\ttfamily arXiv:1406.0755}}\relax
\mciteBstWouldAddEndPuncttrue
\mciteSetBstMidEndSepPunct{\mcitedefaultmidpunct}
{\mcitedefaultendpunct}{\mcitedefaultseppunct}\relax
\EndOfBibitem
\bibitem{LHCb-PAPER-2016-002}
LHCb collaboration, R.~Aaij {\em et~al.}, \ifthenelse{\boolean{articletitles}}{\emph{{Observation of the \mbox{\decay{\Lb}{\Lz\phi}} decay}}, }{}\href{https://doi.org/10.1016/j.physletb.2016.05.077}{Phys.\ Lett.\  \textbf{B759} (2016) 282}, \href{http://arxiv.org/abs/1603.02870}{{\normalfont\ttfamily arXiv:1603.02870}}\relax
\mciteBstWouldAddEndPuncttrue
\mciteSetBstMidEndSepPunct{\mcitedefaultmidpunct}
{\mcitedefaultendpunct}{\mcitedefaultseppunct}\relax
\EndOfBibitem
\bibitem{LHCb-PAPER-2016-004}
LHCb collaboration, R.~Aaij {\em et~al.}, \ifthenelse{\boolean{articletitles}}{\emph{{Observations of \mbox{\decay{\Lb}{\Lz\Kp\pim}} and \mbox{\decay{\Lb}{\Lz\Kp\Km}} decays and searches for other \Lb and \Xibz decays to $\Lz h^+h^-$ final states}}, }{}\href{https://doi.org/10.1007/JHEP05(2016)081}{JHEP \textbf{05} (2016) 081}, \href{http://arxiv.org/abs/1603.00413}{{\normalfont\ttfamily arXiv:1603.00413}}\relax
\mciteBstWouldAddEndPuncttrue
\mciteSetBstMidEndSepPunct{\mcitedefaultmidpunct}
{\mcitedefaultendpunct}{\mcitedefaultseppunct}\relax
\EndOfBibitem
\bibitem{LHCb-PAPER-2016-030}
LHCb collaboration, R.~Aaij {\em et~al.}, \ifthenelse{\boolean{articletitles}}{\emph{{Measurement of matter-antimatter differences in beauty baryon decays}}, }{}\href{https://doi.org/10.1038/nphys4021}{Nature Physics \textbf{13} (2017) 391}, \href{http://arxiv.org/abs/1609.05216}{{\normalfont\ttfamily arXiv:1609.05216}}\relax
\mciteBstWouldAddEndPuncttrue
\mciteSetBstMidEndSepPunct{\mcitedefaultmidpunct}
{\mcitedefaultendpunct}{\mcitedefaultseppunct}\relax
\EndOfBibitem
\bibitem{LHCb-PAPER-2019-028}
LHCb collaboration, R.~Aaij {\em et~al.}, \ifthenelse{\boolean{articletitles}}{\emph{{Search for \CP violation and observation of $P$ violation in \decay{\Lb}{p\pim\pip\pim} decays}}, }{}\href{https://doi.org/10.1103/PhysRevD.102.051101}{Phys.\ Rev.\  \textbf{D102} (2020) 051101}, \href{http://arxiv.org/abs/1912.10741}{{\normalfont\ttfamily arXiv:1912.10741}}\relax
\mciteBstWouldAddEndPuncttrue
\mciteSetBstMidEndSepPunct{\mcitedefaultmidpunct}
{\mcitedefaultendpunct}{\mcitedefaultseppunct}\relax
\EndOfBibitem
\bibitem{LHCb-PAPER-2016-059}
LHCb collaboration, R.~Aaij {\em et~al.}, \ifthenelse{\boolean{articletitles}}{\emph{{Observation of the decay \mbox{\decay{\Lb}{\proton\Km\mumu}} and search for \CP violation}}, }{}\href{https://doi.org/10.1007/JHEP06(2017)108}{JHEP \textbf{06} (2017) 108}, \href{http://arxiv.org/abs/1703.00256}{{\normalfont\ttfamily arXiv:1703.00256}}\relax
\mciteBstWouldAddEndPuncttrue
\mciteSetBstMidEndSepPunct{\mcitedefaultmidpunct}
{\mcitedefaultendpunct}{\mcitedefaultseppunct}\relax
\EndOfBibitem
\bibitem{LHCb-PAPER-2017-044}
LHCb collaboration, R.~Aaij {\em et~al.}, \ifthenelse{\boolean{articletitles}}{\emph{{Search for \CP violation in \mbox{\decay{\Lc}{p \Km \Kp}} and \mbox{\decay{\Lc}{p\pim\pip}} decays}}, }{}\href{https://doi.org/10.1007/JHEP03(2018)182}{JHEP \textbf{03} (2018) 182}, \href{http://arxiv.org/abs/1712.07051}{{\normalfont\ttfamily arXiv:1712.07051}}\relax
\mciteBstWouldAddEndPuncttrue
\mciteSetBstMidEndSepPunct{\mcitedefaultmidpunct}
{\mcitedefaultendpunct}{\mcitedefaultseppunct}\relax
\EndOfBibitem
\bibitem{LHCb-PAPER-2018-001}
LHCb collaboration, R.~Aaij {\em et~al.}, \ifthenelse{\boolean{articletitles}}{\emph{{Search for \CP violation using triple product asymmetries in \mbox{\decay{\Lb}{p\Km\pip\pim}}, \mbox{\decay{\Lb}{p\Km\Kp\Km}}, and \mbox{\decay{\Xires^0_b}{p\Km\Km\pip}} decays}}, }{}\href{https://doi.org/10.1007/JHEP08(2018)039}{JHEP \textbf{08} (2018) 039}, \href{http://arxiv.org/abs/1805.03941}{{\normalfont\ttfamily arXiv:1805.03941}}\relax
\mciteBstWouldAddEndPuncttrue
\mciteSetBstMidEndSepPunct{\mcitedefaultmidpunct}
{\mcitedefaultendpunct}{\mcitedefaultseppunct}\relax
\EndOfBibitem
\bibitem{PhysRevLett.106.181802}
CDF collaboration, T.~Aaltonen {\em et~al.}, \ifthenelse{\boolean{articletitles}}{\emph{{Measurements of direct \CP violating asymmetries in charmless decays of strange bottom mesons and bottom baryons}}, }{}\href{https://doi.org/10.1103/PhysRevLett.106.181802}{Phys.\ Rev.\ Lett.\  \textbf{106} (2011) 181802}, \href{http://arxiv.org/abs/1103.5762}{{\normalfont\ttfamily arXiv:1103.5762}}\relax
\mciteBstWouldAddEndPuncttrue
\mciteSetBstMidEndSepPunct{\mcitedefaultmidpunct}
{\mcitedefaultendpunct}{\mcitedefaultseppunct}\relax
\EndOfBibitem
\bibitem{PhysRevLett.113.242001}
CDF collaboration, T.~Aaltonen {\em et~al.}, \ifthenelse{\boolean{articletitles}}{\emph{{Measurements of direct \CP-violating asymmetries in charmless decays of bottom baryons}}, }{}\href{https://doi.org/10.1103/PhysRevLett.113.242001}{Phys.\ Rev.\ Lett.\  \textbf{113} (2014) 242001}, \href{http://arxiv.org/abs/1403.5586}{{\normalfont\ttfamily arXiv:1403.5586}}\relax
\mciteBstWouldAddEndPuncttrue
\mciteSetBstMidEndSepPunct{\mcitedefaultmidpunct}
{\mcitedefaultendpunct}{\mcitedefaultseppunct}\relax
\EndOfBibitem
\bibitem{LHCb-PAPER-2018-025}
LHCb collaboration, R.~Aaij {\em et~al.}, \ifthenelse{\boolean{articletitles}}{\emph{{Search for \CP violation in \mbox{\decay{\Lb}{p\Km} and \mbox{\decay{\Lb}{p\pim}}} decays}}, }{}\href{https://doi.org/10.1016/j.physletb.2018.10.039}{Phys.\ Lett.\  \textbf{B784} (2018) 101}, \href{http://arxiv.org/abs/1807.06544}{{\normalfont\ttfamily arXiv:1807.06544}}\relax
\mciteBstWouldAddEndPuncttrue
\mciteSetBstMidEndSepPunct{\mcitedefaultmidpunct}
{\mcitedefaultendpunct}{\mcitedefaultseppunct}\relax
\EndOfBibitem
\bibitem{LHCb-PAPER-2018-044}
LHCb collaboration, R.~Aaij {\em et~al.}, \ifthenelse{\boolean{articletitles}}{\emph{{Measurement of \CP asymmetries in charmless four-body \Lb and \Xibz decays}}, }{}\href{https://doi.org/10.1140/epjc/s10052-019-7218-1}{Eur.\ Phys.\ J.\  \textbf{C79} (2019) 745}, \href{http://arxiv.org/abs/1903.06792}{{\normalfont\ttfamily arXiv:1903.06792}}\relax
\mciteBstWouldAddEndPuncttrue
\mciteSetBstMidEndSepPunct{\mcitedefaultmidpunct}
{\mcitedefaultendpunct}{\mcitedefaultseppunct}\relax
\EndOfBibitem
\bibitem{LHCb-PAPER-2019-026}
LHCb collaboration, R.~Aaij {\em et~al.}, \ifthenelse{\boolean{articletitles}}{\emph{{Search for \CP violation in \mbox{\decay{\Xicp}{p\Km\pip}} decays with model-independent techniques}}, }{}\href{https://doi.org/10.1140/epjc/s10052-020-8365-0}{Eur.\ Phys.\ J.\  \textbf{C80} (2020) 986}, \href{http://arxiv.org/abs/2006.03145}{{\normalfont\ttfamily arXiv:2006.03145}}\relax
\mciteBstWouldAddEndPuncttrue
\mciteSetBstMidEndSepPunct{\mcitedefaultmidpunct}
{\mcitedefaultendpunct}{\mcitedefaultseppunct}\relax
\EndOfBibitem
\bibitem{LHCb-PAPER-2020-017}
LHCb collaboration, R.~Aaij {\em et~al.}, \ifthenelse{\boolean{articletitles}}{\emph{{Search for \CP violation in $\Xibm \to p K^- K^- decays$}}, }{}\href{https://doi.org/10.1103/PhysRevD.104.052010}{Phys.\ Rev.\  \textbf{D104} (2021) 052010}, \href{http://arxiv.org/abs/2104.15074}{{\normalfont\ttfamily arXiv:2104.15074}}\relax
\mciteBstWouldAddEndPuncttrue
\mciteSetBstMidEndSepPunct{\mcitedefaultmidpunct}
{\mcitedefaultendpunct}{\mcitedefaultseppunct}\relax
\EndOfBibitem
\bibitem{LHCb-PAPER-2021-027}
LHCb collaboration, R.~Aaij {\em et~al.}, \ifthenelse{\boolean{articletitles}}{\emph{{Observation of the suppressed $\Lb \to \D \Pp \Km$ decay with $\D \to \Kp \pim$ and measurement of its \CP asymmetry}}, }{}\href{https://doi.org/10.1103/PhysRevD.104.112008}{Phys.\ Rev.\  \textbf{D104} (2021) 112008}, \href{http://arxiv.org/abs/2109.02621}{{\normalfont\ttfamily arXiv:2109.02621}}\relax
\mciteBstWouldAddEndPuncttrue
\mciteSetBstMidEndSepPunct{\mcitedefaultmidpunct}
{\mcitedefaultendpunct}{\mcitedefaultseppunct}\relax
\EndOfBibitem
\bibitem{LHCb-PAPER-2021-030}
LHCb collaboration, R.~Aaij {\em et~al.}, \ifthenelse{\boolean{articletitles}}{\emph{{Measurement of the photon polarization in $\Lb \to \Lz \gamma$ decays}}, }{}\href{https://doi.org/10.1103/PhysRevD.105.L051104}{Phys.\ Rev.\  \textbf{D105} (2022) 051104}, \href{http://arxiv.org/abs/2111.10194}{{\normalfont\ttfamily arXiv:2111.10194}}\relax
\mciteBstWouldAddEndPuncttrue
\mciteSetBstMidEndSepPunct{\mcitedefaultmidpunct}
{\mcitedefaultendpunct}{\mcitedefaultseppunct}\relax
\EndOfBibitem
\bibitem{LHCb-PAPER-2024-048}
LHCb collaboration, R.~Aaij {\em et~al.}, \ifthenelse{\boolean{articletitles}}{\emph{{Measurement of $C\!P$ asymmetries in $\Lb \to ph^-$ decays}}, }{}\href{https://doi.org/10.1103/PhysRevD.111.092004}{{Phys.\ Rev.\ } \textbf{D111} (2025) 092004}, \href{http://arxiv.org/abs/2412.13958}{{\normalfont\ttfamily arXiv:2412.13958}}\relax
\mciteBstWouldAddEndPuncttrue
\mciteSetBstMidEndSepPunct{\mcitedefaultmidpunct}
{\mcitedefaultendpunct}{\mcitedefaultseppunct}\relax
\EndOfBibitem
\bibitem{LHCb-PAPER-2024-054}
LHCb collaboration, R.~Aaij {\em et~al.}, \ifthenelse{\boolean{articletitles}}{\emph{{Observation of charge-parity symmetry breaking in baryon decays}}, }{}\href{https://doi.org/10.1038/s41586-025-09119-3}{{Nature} \textbf{643} (2025) 1223}, \href{http://arxiv.org/abs/2503.16954}{{\normalfont\ttfamily arXiv:2503.16954}}\relax
\mciteBstWouldAddEndPuncttrue
\mciteSetBstMidEndSepPunct{\mcitedefaultmidpunct}
{\mcitedefaultendpunct}{\mcitedefaultseppunct}\relax
\EndOfBibitem
\bibitem{LHCb-PAPER-2024-043}
LHCb collaboration, R.~Aaij {\em et~al.}, \ifthenelse{\boolean{articletitles}}{\emph{{Study of $\Lambda^{0}_{b}$ and $\Xi^{0}_{b}$ decays to $\Lambda h^+ h^{\prime -}$ and evidence for \CP violation in $\Lambda^{0}_{b} \to \Lambda K^+ K^-$}}, }{}\href{https://doi.org/10.1103/PhysRevLett.134.101802}{{Phys.\ Rev.\ Lett.\ } \textbf{134} (2025) 101802}, \href{http://arxiv.org/abs/2411.15441}{{\normalfont\ttfamily arXiv:2411.15441}}\relax
\mciteBstWouldAddEndPuncttrue
\mciteSetBstMidEndSepPunct{\mcitedefaultmidpunct}
{\mcitedefaultendpunct}{\mcitedefaultseppunct}\relax
\EndOfBibitem
\bibitem{PhysRevD.91.116007}
Y.~K. Hsiao and C.~Q. Geng, \ifthenelse{\boolean{articletitles}}{\emph{{Direct $\CP$ violation in $\Lb$ decays}}, }{}\href{https://doi.org/10.1103/PhysRevD.91.116007}{Phys.\ Rev.\  \textbf{D91} (2015) 116007}, \href{http://arxiv.org/abs/1412.1899}{{\normalfont\ttfamily arXiv:1412.1899}}\relax
\mciteBstWouldAddEndPuncttrue
\mciteSetBstMidEndSepPunct{\mcitedefaultmidpunct}
{\mcitedefaultendpunct}{\mcitedefaultseppunct}\relax
\EndOfBibitem
\bibitem{PhysRevD.95.093001}
Y.~K. Hsiao, Y.~Yao, and C.~Q. Geng, \ifthenelse{\boolean{articletitles}}{\emph{Charmless two-body antitriplet $b$-baryon decays}, }{}\href{https://doi.org/10.1103/PhysRevD.95.093001}{Phys.\ Rev.\  \textbf{D95} (2017) 093001}, \href{http://arxiv.org/abs/1702.05263}{{\normalfont\ttfamily arXiv:1702.05263}}\relax
\mciteBstWouldAddEndPuncttrue
\mciteSetBstMidEndSepPunct{\mcitedefaultmidpunct}
{\mcitedefaultendpunct}{\mcitedefaultseppunct}\relax
\EndOfBibitem
\bibitem{PhysRevD.99.054020}
J.~Zhu, Z.-T. Wei, and H.-W. Ke, \ifthenelse{\boolean{articletitles}}{\emph{{Semileptonic and nonleptonic weak decays of $\Lb$}}, }{}\href{https://doi.org/10.1103/PhysRevD.99.054020}{Phys.\ Rev.\  \textbf{D99} (2019) 054020}, \href{http://arxiv.org/abs/1803.01297}{{\normalfont\ttfamily arXiv:1803.01297}}\relax
\mciteBstWouldAddEndPuncttrue
\mciteSetBstMidEndSepPunct{\mcitedefaultmidpunct}
{\mcitedefaultendpunct}{\mcitedefaultseppunct}\relax
\EndOfBibitem
\bibitem{GENG2021136125}
C.~Q. Geng, C.-W. Liu, and T.-H. Tsai, \ifthenelse{\boolean{articletitles}}{\emph{{Non-leptonic two-body decays of \Lb in light-front quark model}}, }{}\href{https://doi.org/https://doi.org/10.1016/j.physletb.2021.136125}{Phys.\ Lett.\  \textbf{B815} (2021) 136125}, \href{http://arxiv.org/abs/2102.01552}{{\normalfont\ttfamily arXiv:2102.01552}}\relax
\mciteBstWouldAddEndPuncttrue
\mciteSetBstMidEndSepPunct{\mcitedefaultmidpunct}
{\mcitedefaultendpunct}{\mcitedefaultseppunct}\relax
\EndOfBibitem
\bibitem{han2025establishingcpviolationbbaryon}
J.-J. Han {\em et~al.}, \ifthenelse{\boolean{articletitles}}{\emph{{Establishing \CP violation in b-baryon decays}}, }{}\href{https://doi.org/10.1103/ynnx-f63h}{Phys.\ Rev.\ Lett.\  \textbf{134} (2025) 221801}, \href{http://arxiv.org/abs/2409.02821}{{\normalfont\ttfamily arXiv:2409.02821}}\relax
\mciteBstWouldAddEndPuncttrue
\mciteSetBstMidEndSepPunct{\mcitedefaultmidpunct}
{\mcitedefaultendpunct}{\mcitedefaultseppunct}\relax
\EndOfBibitem
\bibitem{duan2024finalstaterescatteringmechanismbottombaryon}
Z.-D. Duan {\em et~al.}, \ifthenelse{\boolean{articletitles}}{\emph{Final-state rescattering mechanism of bottom-baryon decays}, }{}\href{http://arxiv.org/abs/2412.20458}{{\normalfont\ttfamily arXiv:2412.20458}}\relax
\mciteBstWouldAddEndPuncttrue
\mciteSetBstMidEndSepPunct{\mcitedefaultmidpunct}
{\mcitedefaultendpunct}{\mcitedefaultseppunct}\relax
\EndOfBibitem
\bibitem{Dalitz:1953cp}
R.~H. Dalitz, \ifthenelse{\boolean{articletitles}}{\emph{{On the analysis of \Ptau-meson data and the nature of the \Ptau-meson}}, }{}\href{https://doi.org/10.1080/14786441008520365}{Phil.\ Mag.\ Ser.\ 7 \textbf{44} (1953) 1068}\relax
\mciteBstWouldAddEndPuncttrue
\mciteSetBstMidEndSepPunct{\mcitedefaultmidpunct}
{\mcitedefaultendpunct}{\mcitedefaultseppunct}\relax
\EndOfBibitem
\bibitem{Fabri:1954zz}
E.~Fabri, \ifthenelse{\boolean{articletitles}}{\emph{{A study of \Ptau-meson decay}}, }{}\href{https://doi.org/10.1007/BF02781042}{Nuovo Cim.\  \textbf{11} (1954) 479}\relax
\mciteBstWouldAddEndPuncttrue
\mciteSetBstMidEndSepPunct{\mcitedefaultmidpunct}
{\mcitedefaultendpunct}{\mcitedefaultseppunct}\relax
\EndOfBibitem
\bibitem{LHCb-DP-2008-001}
LHCb collaboration, A.~A. Alves~Jr.\ {\em et~al.}, \ifthenelse{\boolean{articletitles}}{\emph{{The \lhcb detector at the LHC}}, }{}\href{https://doi.org/10.1088/1748-0221/3/08/S08005}{JINST \textbf{3} (2008) S08005}\relax
\mciteBstWouldAddEndPuncttrue
\mciteSetBstMidEndSepPunct{\mcitedefaultmidpunct}
{\mcitedefaultendpunct}{\mcitedefaultseppunct}\relax
\EndOfBibitem
\bibitem{LHCb-DP-2014-002}
LHCb collaboration, R.~Aaij {\em et~al.}, \ifthenelse{\boolean{articletitles}}{\emph{{LHCb detector performance}}, }{}\href{https://doi.org/10.1142/S0217751X15300227}{Int.\ J.\ Mod.\ Phys.\  \textbf{A30} (2015) 1530022}, \href{http://arxiv.org/abs/1412.6352}{{\normalfont\ttfamily arXiv:1412.6352}}\relax
\mciteBstWouldAddEndPuncttrue
\mciteSetBstMidEndSepPunct{\mcitedefaultmidpunct}
{\mcitedefaultendpunct}{\mcitedefaultseppunct}\relax
\EndOfBibitem
\bibitem{BBDT}
V.~V. Gligorov and M.~Williams, \ifthenelse{\boolean{articletitles}}{\emph{{Efficient, reliable and fast high-level triggering using a bonsai boosted decision tree}}, }{}\href{https://doi.org/10.1088/1748-0221/8/02/P02013}{JINST \textbf{8} (2013) P02013}, \href{http://arxiv.org/abs/1210.6861}{{\normalfont\ttfamily arXiv:1210.6861}}\relax
\mciteBstWouldAddEndPuncttrue
\mciteSetBstMidEndSepPunct{\mcitedefaultmidpunct}
{\mcitedefaultendpunct}{\mcitedefaultseppunct}\relax
\EndOfBibitem
\bibitem{LHCb-PROC-2015-018}
T.~Likhomanenko {\em et~al.}, \ifthenelse{\boolean{articletitles}}{\emph{{LHCb topological trigger reoptimization}}, }{}\href{https://doi.org/10.1088/1742-6596/664/8/082025}{J.\ Phys.\ Conf.\ Ser.\  \textbf{664} (2015) 082025}, \href{http://arxiv.org/abs/1510.00572}{{\normalfont\ttfamily arXiv:1510.00572}}\relax
\mciteBstWouldAddEndPuncttrue
\mciteSetBstMidEndSepPunct{\mcitedefaultmidpunct}
{\mcitedefaultendpunct}{\mcitedefaultseppunct}\relax
\EndOfBibitem
\bibitem{Sjostrand:2007gs}
T.~Sj\"{o}strand, S.~Mrenna, and P.~Skands, \ifthenelse{\boolean{articletitles}}{\emph{{A brief introduction to PYTHIA 8.1}}, }{}\href{https://doi.org/10.1016/j.cpc.2008.01.036}{Comput.\ Phys.\ Commun.\  \textbf{178} (2008) 852}, \href{http://arxiv.org/abs/0710.3820}{{\normalfont\ttfamily arXiv:0710.3820}}\relax
\mciteBstWouldAddEndPuncttrue
\mciteSetBstMidEndSepPunct{\mcitedefaultmidpunct}
{\mcitedefaultendpunct}{\mcitedefaultseppunct}\relax
\EndOfBibitem
\bibitem{Sjostrand:2006za}
T.~Sj\"{o}strand, S.~Mrenna, and P.~Skands, \ifthenelse{\boolean{articletitles}}{\emph{{PYTHIA 6.4 physics and manual}}, }{}\href{https://doi.org/10.1088/1126-6708/2006/05/026}{JHEP \textbf{05} (2006) 026}, \href{http://arxiv.org/abs/hep-ph/0603175}{{\normalfont\ttfamily arXiv:hep-ph/0603175}}\relax
\mciteBstWouldAddEndPuncttrue
\mciteSetBstMidEndSepPunct{\mcitedefaultmidpunct}
{\mcitedefaultendpunct}{\mcitedefaultseppunct}\relax
\EndOfBibitem
\bibitem{LHCb-PROC-2010-056}
I.~Belyaev {\em et~al.}, \ifthenelse{\boolean{articletitles}}{\emph{{Handling of the generation of primary events in Gauss, the LHCb simulation framework}}, }{}\href{https://doi.org/10.1088/1742-6596/331/3/032047}{J.\ Phys.\ Conf.\ Ser.\  \textbf{331} (2011) 032047}\relax
\mciteBstWouldAddEndPuncttrue
\mciteSetBstMidEndSepPunct{\mcitedefaultmidpunct}
{\mcitedefaultendpunct}{\mcitedefaultseppunct}\relax
\EndOfBibitem
\bibitem{Lange:2001uf}
D.~J. Lange, \ifthenelse{\boolean{articletitles}}{\emph{{The EvtGen particle decay simulation package}}, }{}\href{https://doi.org/10.1016/S0168-9002(01)00089-4}{Nucl.\ Instrum.\ Meth.\  \textbf{A462} (2001) 152}\relax
\mciteBstWouldAddEndPuncttrue
\mciteSetBstMidEndSepPunct{\mcitedefaultmidpunct}
{\mcitedefaultendpunct}{\mcitedefaultseppunct}\relax
\EndOfBibitem
\bibitem{davidson2015photos}
N.~Davidson, T.~Przedzinski, and Z.~Was, \ifthenelse{\boolean{articletitles}}{\emph{{PHOTOS interface in C++: Technical and physics documentation}}, }{}\href{https://doi.org/https://doi.org/10.1016/j.cpc.2015.09.013}{Comput.\ Phys.\ Commun.\  \textbf{199} (2016) 86}, \href{http://arxiv.org/abs/1011.0937}{{\normalfont\ttfamily arXiv:1011.0937}}\relax
\mciteBstWouldAddEndPuncttrue
\mciteSetBstMidEndSepPunct{\mcitedefaultmidpunct}
{\mcitedefaultendpunct}{\mcitedefaultseppunct}\relax
\EndOfBibitem
\bibitem{Allison:2006ve}
Geant4 collaboration, J.~Allison {\em et~al.}, \ifthenelse{\boolean{articletitles}}{\emph{{Geant4 developments and applications}}, }{}\href{https://doi.org/10.1109/TNS.2006.869826}{IEEE Trans.\ Nucl.\ Sci.\  \textbf{53} (2006) 270}\relax
\mciteBstWouldAddEndPuncttrue
\mciteSetBstMidEndSepPunct{\mcitedefaultmidpunct}
{\mcitedefaultendpunct}{\mcitedefaultseppunct}\relax
\EndOfBibitem
\bibitem{Agostinelli:2002hh}
Geant4 collaboration, S.~Agostinelli {\em et~al.}, \ifthenelse{\boolean{articletitles}}{\emph{{Geant4: A simulation toolkit}}, }{}\href{https://doi.org/10.1016/S0168-9002(03)01368-8}{Nucl.\ Instrum.\ Meth.\  \textbf{A506} (2003) 250}\relax
\mciteBstWouldAddEndPuncttrue
\mciteSetBstMidEndSepPunct{\mcitedefaultmidpunct}
{\mcitedefaultendpunct}{\mcitedefaultseppunct}\relax
\EndOfBibitem
\bibitem{LHCb-PROC-2011-006}
M.~Clemencic {\em et~al.}, \ifthenelse{\boolean{articletitles}}{\emph{{The \lhcb simulation application, Gauss: Design, evolution and experience}}, }{}\href{https://doi.org/10.1088/1742-6596/331/3/032023}{J.\ Phys.\ Conf.\ Ser.\  \textbf{331} (2011) 032023}\relax
\mciteBstWouldAddEndPuncttrue
\mciteSetBstMidEndSepPunct{\mcitedefaultmidpunct}
{\mcitedefaultendpunct}{\mcitedefaultseppunct}\relax
\EndOfBibitem
\bibitem{PDG2024}
Particle Data Group, S.~Navas {\em et~al.}, \ifthenelse{\boolean{articletitles}}{\emph{{\href{http://pdg.lbl.gov/}{Review of particle physics}}}, }{}\href{https://doi.org/10.1103/PhysRevD.110.030001}{Phys.\ Rev.\  \textbf{D110} (2024) 030001}\relax
\mciteBstWouldAddEndPuncttrue
\mciteSetBstMidEndSepPunct{\mcitedefaultmidpunct}
{\mcitedefaultendpunct}{\mcitedefaultseppunct}\relax
\EndOfBibitem
\bibitem{Breiman}
L.~Breiman, J.~H. Friedman, R.~A. Olshen, and C.~J. Stone, {\em Classification and regression trees}, Wadsworth international group, Belmont, California, USA, 1984\relax
\mciteBstWouldAddEndPuncttrue
\mciteSetBstMidEndSepPunct{\mcitedefaultmidpunct}
{\mcitedefaultendpunct}{\mcitedefaultseppunct}\relax
\EndOfBibitem
\bibitem{AdaBoost}
Y.~Freund and R.~E. Schapire, \ifthenelse{\boolean{articletitles}}{\emph{A decision-theoretic generalization of on-line learning and an application to boosting}, }{}\href{https://doi.org/10.1006/jcss.1997.1504}{J.\ Comput.\ Syst.\ Sci.\  \textbf{55} (1997) 119}\relax
\mciteBstWouldAddEndPuncttrue
\mciteSetBstMidEndSepPunct{\mcitedefaultmidpunct}
{\mcitedefaultendpunct}{\mcitedefaultseppunct}\relax
\EndOfBibitem
\bibitem{Punzi:2003bu}
G.~Punzi, \ifthenelse{\boolean{articletitles}}{\emph{{Sensitivity of searches for new signals and its optimization}}, }{}eConf \textbf{C030908} (2003) MODT002, \href{http://arxiv.org/abs/physics/0308063}{{\normalfont\ttfamily arXiv:physics/0308063}}\relax
\mciteBstWouldAddEndPuncttrue
\mciteSetBstMidEndSepPunct{\mcitedefaultmidpunct}
{\mcitedefaultendpunct}{\mcitedefaultseppunct}\relax
\EndOfBibitem
\bibitem{LHCb-PAPER-2018-047}
LHCb collaboration, R.~Aaij {\em et~al.}, \ifthenelse{\boolean{articletitles}}{\emph{{Measurement of the mass and production rate of \Xibm baryons}}, }{}\href{https://doi.org/10.1103/PhysRevD.99.052006}{Phys.\ Rev.\  \textbf{D99} (2019) 052006}, \href{http://arxiv.org/abs/1901.07075}{{\normalfont\ttfamily arXiv:1901.07075}}\relax
\mciteBstWouldAddEndPuncttrue
\mciteSetBstMidEndSepPunct{\mcitedefaultmidpunct}
{\mcitedefaultendpunct}{\mcitedefaultseppunct}\relax
\EndOfBibitem
\bibitem{Skwarnicki:1986xj}
T.~Skwarnicki, {\em {A study of the radiative cascade transitions between the Upsilon-prime and Upsilon resonances}}, PhD thesis, Institute of Nuclear Physics, Krakow, 1986, {\href{http://inspirehep.net/record/230779/}{DESY-F31-86-02}}\relax
\mciteBstWouldAddEndPuncttrue
\mciteSetBstMidEndSepPunct{\mcitedefaultmidpunct}
{\mcitedefaultendpunct}{\mcitedefaultseppunct}\relax
\EndOfBibitem
\bibitem{ARGUS:1990hfq}
ARGUS collaboration, H.~Albrecht {\em et~al.}, \ifthenelse{\boolean{articletitles}}{\emph{{Search for hadronic $b \to u$ decays}}, }{}\href{https://doi.org/10.1016/0370-2693(90)91293-K}{Phys.\ Lett.\  \textbf{B241} (1990) 278}\relax
\mciteBstWouldAddEndPuncttrue
\mciteSetBstMidEndSepPunct{\mcitedefaultmidpunct}
{\mcitedefaultendpunct}{\mcitedefaultseppunct}\relax
\EndOfBibitem
\bibitem{Wilks:1938dza}
S.~S. Wilks, \ifthenelse{\boolean{articletitles}}{\emph{{The large-sample distribution of the likelihood ratio for testing composite hypotheses}}, }{}\href{https://doi.org/10.1214/aoms/1177732360}{Ann.\ Math.\ Stat.\  \textbf{9} (1938) 60}\relax
\mciteBstWouldAddEndPuncttrue
\mciteSetBstMidEndSepPunct{\mcitedefaultmidpunct}
{\mcitedefaultendpunct}{\mcitedefaultseppunct}\relax
\EndOfBibitem
\bibitem{LHCb-DP-2013-002}
LHCb collaboration, R.~Aaij {\em et~al.}, \ifthenelse{\boolean{articletitles}}{\emph{{Measurement of the track reconstruction efficiency at LHCb}}, }{}\href{https://doi.org/10.1088/1748-0221/10/02/P02007}{JINST \textbf{10} (2015) P02007}, \href{http://arxiv.org/abs/1408.1251}{{\normalfont\ttfamily arXiv:1408.1251}}\relax
\mciteBstWouldAddEndPuncttrue
\mciteSetBstMidEndSepPunct{\mcitedefaultmidpunct}
{\mcitedefaultendpunct}{\mcitedefaultseppunct}\relax
\EndOfBibitem
\bibitem{LHCb-PUB-2016-021}
L.~Anderlini {\em et~al.}, \ifthenelse{\boolean{articletitles}}{\emph{{The PIDCalib package}}, }{} \href{http://cdsweb.cern.ch/search?p=LHCb-PUB-2016-021&f=reportnumber&action_search=Search&c=LHCb+Notes} {LHCb-PUB-2016-021}, 2016\relax
\mciteBstWouldAddEndPuncttrue
\mciteSetBstMidEndSepPunct{\mcitedefaultmidpunct}
{\mcitedefaultendpunct}{\mcitedefaultseppunct}\relax
\EndOfBibitem
\bibitem{LHCb-DP-2018-001}
R.~Aaij {\em et~al.}, \ifthenelse{\boolean{articletitles}}{\emph{{Selection and processing of calibration samples to measure the particle identification performance of the LHCb experiment in Run 2}}, }{}\href{https://doi.org/10.1140/epjti/s40485-019-0050-z}{Eur.\ Phys.\ J.\ Tech.\ Instr.\  \textbf{6} (2019) 1}, \href{http://arxiv.org/abs/1803.00824}{{\normalfont\ttfamily arXiv:1803.00824}}\relax
\mciteBstWouldAddEndPuncttrue
\mciteSetBstMidEndSepPunct{\mcitedefaultmidpunct}
{\mcitedefaultendpunct}{\mcitedefaultseppunct}\relax
\EndOfBibitem
\bibitem{LHCb-DP-2012-004}
R.~Aaij {\em et~al.}, \ifthenelse{\boolean{articletitles}}{\emph{{The \lhcb trigger and its performance in 2011}}, }{}\href{https://doi.org/10.1088/1748-0221/8/04/P04022}{JINST \textbf{8} (2013) P04022}, \href{http://arxiv.org/abs/1211.3055}{{\normalfont\ttfamily arXiv:1211.3055}}\relax
\mciteBstWouldAddEndPuncttrue
\mciteSetBstMidEndSepPunct{\mcitedefaultmidpunct}
{\mcitedefaultendpunct}{\mcitedefaultseppunct}\relax
\EndOfBibitem
\bibitem{Rogozhnikov:2016bdp}
A.~Rogozhnikov, \ifthenelse{\boolean{articletitles}}{\emph{{Reweighting with boosted decision trees}}, }{}\href{https://doi.org/10.1088/1742-6596/762/1/012036}{J.\ Phys.\ Conf.\ Ser.\  \textbf{762} (2016) 012036}, \href{http://arxiv.org/abs/1608.05806}{{\normalfont\ttfamily arXiv:1608.05806}}, \url{https://github.com/arogozhnikov/hep_ml}\relax
\mciteBstWouldAddEndPuncttrue
\mciteSetBstMidEndSepPunct{\mcitedefaultmidpunct}
{\mcitedefaultendpunct}{\mcitedefaultseppunct}\relax
\EndOfBibitem
\bibitem{LHCb-PAPER-2012-057}
LHCb collaboration, R.~Aaij {\em et~al.}, \ifthenelse{\boolean{articletitles}}{\emph{{Measurements of the \mbox{\decay{\Lb}{\jpsi\Lz}} decay amplitudes and the \Lb polarisation in \proton\proton collisions at $\sqs = $7~\tev}}, }{}\href{https://doi.org/10.1016/j.physletb.2013.05.041}{Phys.\ Lett.\  \textbf{B724} (2013) 27}, \href{http://arxiv.org/abs/1302.5578}{{\normalfont\ttfamily arXiv:1302.5578}}\relax
\mciteBstWouldAddEndPuncttrue
\mciteSetBstMidEndSepPunct{\mcitedefaultmidpunct}
{\mcitedefaultendpunct}{\mcitedefaultseppunct}\relax
\EndOfBibitem
\bibitem{LHCb-PAPER-2020-005}
LHCb collaboration, R.~Aaij {\em et~al.}, \ifthenelse{\boolean{articletitles}}{\emph{{Measurement of the $\Lb \to \jpsi \Lz$ angular distribution and the \Lz polarisation in $pp$ collisions}}, }{}\href{https://doi.org/10.1007/JHEP06(2020)110}{JHEP \textbf{06} (2020) 110}, \href{http://arxiv.org/abs/2004.10563}{{\normalfont\ttfamily arXiv:2004.10563}}\relax
\mciteBstWouldAddEndPuncttrue
\mciteSetBstMidEndSepPunct{\mcitedefaultmidpunct}
{\mcitedefaultendpunct}{\mcitedefaultseppunct}\relax
\EndOfBibitem
\bibitem{LHCb-PAPER-2024-017}
LHCb collaboration, R.~Aaij {\em et~al.}, \ifthenelse{\boolean{articletitles}}{\emph{{Measurement of $\Lambda_b^0$, $\Lambda_c^+$ and $\Lambda$ decay parameters using $\Lambda_b^0 \to \Lambda_c^+h^-$ decays}}, }{}\href{https://doi.org/10.1103/PhysRevLett.133.261804}{{Phys.\ Rev.\ Lett.\ } \textbf{133} (2024) 261804}, \href{http://arxiv.org/abs/2409.02759}{{\normalfont\ttfamily arXiv:2409.02759}}\relax
\mciteBstWouldAddEndPuncttrue
\mciteSetBstMidEndSepPunct{\mcitedefaultmidpunct}
{\mcitedefaultendpunct}{\mcitedefaultseppunct}\relax
\EndOfBibitem
\bibitem{Pivk:2004ty}
M.~Pivk and F.~R. Le~Diberder, \ifthenelse{\boolean{articletitles}}{\emph{{sPlot: A statistical tool to unfold data distributions}}, }{}\href{https://doi.org/10.1016/j.nima.2005.08.106}{Nucl.\ Instrum.\ Meth.\  \textbf{A555} (2005) 356}, \href{http://arxiv.org/abs/physics/0402083}{{\normalfont\ttfamily arXiv:physics/0402083}}\relax
\mciteBstWouldAddEndPuncttrue
\mciteSetBstMidEndSepPunct{\mcitedefaultmidpunct}
{\mcitedefaultendpunct}{\mcitedefaultseppunct}\relax
\EndOfBibitem
\bibitem{Santos:2013gra}
D.~Mart{\'\i}nez~Santos and F.~Dupertuis, \ifthenelse{\boolean{articletitles}}{\emph{{Mass distributions marginalized over per-event errors}}, }{}\href{https://doi.org/10.1016/j.nima.2014.06.081}{Nucl.\ Instrum.\ Meth.\  \textbf{A764} (2014) 150}, \href{http://arxiv.org/abs/1312.5000}{{\normalfont\ttfamily arXiv:1312.5000}}\relax
\mciteBstWouldAddEndPuncttrue
\mciteSetBstMidEndSepPunct{\mcitedefaultmidpunct}
{\mcitedefaultendpunct}{\mcitedefaultseppunct}\relax
\EndOfBibitem
\bibitem{CRANMER2001198}
K.~Cranmer, \ifthenelse{\boolean{articletitles}}{\emph{Kernel estimation in high-energy physics}, }{}\href{https://doi.org/https://doi.org/10.1016/S0010-4655(00)00243-5}{Comput.\ Phys.\ Commun.\  \textbf{136} (2001) 198}\relax
\mciteBstWouldAddEndPuncttrue
\mciteSetBstMidEndSepPunct{\mcitedefaultmidpunct}
{\mcitedefaultendpunct}{\mcitedefaultseppunct}\relax
\EndOfBibitem
\bibitem{LHCb-PAPER-2016-062}
LHCb collaboration, R.~Aaij {\em et~al.}, \ifthenelse{\boolean{articletitles}}{\emph{{Measurement of \Bd, \Bs, \Bp and \Lb production asymmetries in 7 and 8~\tev\ proton-proton collisions}}, }{}\href{https://doi.org/10.1016/j.physletb.2017.09.023}{Phys.\ Lett.\  \textbf{B774} (2017) 139}, \href{http://arxiv.org/abs/1703.08464}{{\normalfont\ttfamily arXiv:1703.08464}}\relax
\mciteBstWouldAddEndPuncttrue
\mciteSetBstMidEndSepPunct{\mcitedefaultmidpunct}
{\mcitedefaultendpunct}{\mcitedefaultseppunct}\relax
\EndOfBibitem
\end{mcitethebibliography}
